\documentclass[epsfig,12pt]{article}
\usepackage{epsfig}
\usepackage{graphicx}
\usepackage{rotating}

\usepackage{latexsym}
\usepackage{amsmath}
\usepackage{amssymb}
\usepackage{relsize}
\usepackage{geometry}
\usepackage{color}
\usepackage{bm}
\def\beq{\begin{equation}}
\def\eeq{\end{equation}}
\def\beqn{\begin{eqnarray}}
\def\eeqn{\end{eqnarray}}

\newcommand{\ntwo}{${\mathcal N}=2\;$}

\newcommand{\gsim}{\lower.7ex\hbox{$
\;\stackrel{\textstyle>}{\sim}\;$}}
\newcommand{\lsim}{\lower.7ex\hbox{$
\;\stackrel{\textstyle<}{\sim}\;$}}

\def\beqn{\begin{eqnarray}}
\def\eeqn{\end{eqnarray}}

\def\beq{\begin{equation}}
\def\eeq{\end{equation}}
\def\ba{\beq\new\begin{array}{c}}
\def\ea{\end{array}\eeq}

\newcommand{\ntwot}{${\mathcal N}= \left(2,2\right) $ }

\newcommand{\p}{\partial}

\newcommand{\ov}{\overline}
\newcommand{\mc}[1]{\mathcal{#1}}

\newcommand{\lgr}{\left\lgroup}
\newcommand{\rgr}{\right\rgroup}

\def\slashed#1{\setbox0=\hbox{$#1$}             % set a box for #1
   \dimen0=\wd0                                 % and get its size
   \setbox1=\hbox{/} \dimen1=\wd1               % get size of /
   \ifdim\dimen0>\dimen1                        % #1 is bigger
      \rlap{\hbox to \dimen0{\hfil/\hfil}}      % so center / in box
      #1                                        % and print #1
   \else                                        % / is bigger
      \rlap{\hbox to \dimen1{\hfil$#1$\hfil}}   % so center #1
      /                                         % and print /
   \fi}                                        %

\newcommand{\bpsi}{\ov{\psi}{}}
\newcommand{\bphi}{\ov{\phi}{}}
\newcommand{\bxi}{\ov{\xi}{}}

\newcommand{\bj}{{\bar \jmath}}
\newcommand{\bk}{{\bar k}}

\newcommand{\bmm}{{\bar m}}
\newcommand{\bp}{{\bar p}}
\newcommand{\bkk}{{\bar k}}
\newcommand{\br}{{\bar r}}

\newcommand{\W}{\mathcal{W}}
\newcommand{\M}{\mathcal{M}}
\newcommand{\bM}{\ov{\mathcal{M}}{}}
\newcommand{\Q}{\mathcal{Q}}

\newcommand{\D}{\mathcal{D}}

\newcommand{\V}{\mathcal{V}}

\begin{document}

\hyphenation{con-fi-ning}
\hyphenation{Cou-lomb}
\hyphenation{Yan-ki-e-lo-wicz}
\hyphenation{di-men-si-on-al}
\hyphenation{mo-no-pole}
\hyphenation{me-thod}

%%%%%%%%%%%%%%%%%%%%%%%%%%%%%%%%

\begin{titlepage}

\begin{flushright}
%Draft\\
%
FTPI-MINN-12/08, UMN-TH-3035/12\\
%February 15, 2012
\end{flushright}

\vspace{0.3cm}

\begin{center}
{  \Large \bf  
			2D\,--\,4D Correspondence: Towers of Kinks Versus\\[1mm]  
			Towers of Monopoles
			in \boldmath{${\mathcal N}=2$} Theories
			
}
\end{center}
\vspace{0.6cm}

\begin{center}

 {\large
 \bf   Pavel A.~Bolokhov$^{\,a,b}$,  Mikhail Shifman$^{\,a}$ and Alexei Yung$^{\,\,a,c}$}
\end {center}

\begin{center}

%\vspace{3mm}
$^a${\it  William I. Fine Theoretical Physics Institute, University of Minnesota,
Minneapolis, MN 55455, USA}\\
$^b${\it Theoretical Physics Department, St.Petersburg State University, Ulyanovskaya~1, 
	 Peterhof, St.Petersburg, 198504, Russia}\\
$^{c}${\it Petersburg Nuclear Physics Institute, Gatchina, St. Petersburg
188300, Russia
}

\vspace{0.7cm}

{\large\bf Abstract}
\end{center}

	We continue to study the BPS spectrum of the ${\mathcal N}=(2,2)$ CP$^{N-1}$ model  with 
	the $ \mc{Z}_N $-symmetric twisted mass terms.
	We focus on analysis of the ``extra'' towers found previously in \cite{Bolokhov:2011mp},
	and compare them to the states that can be identified in the quasiclassical domain.
	Exact analysis of the strong-coupling states shows that not all of them survive when passing to the
	weak-coupling domain. 
	Some of the states decay on the curves of the marginal stability (CMS).
	Thus, not all strong-coupling states can be analytically continued to weak coupling 
	to match the observable bound states.
	
	At weak coupling, we confirm the existence of bound states 
	of topologically charged kinks and elementary quanta. 
	Quantization of the U(1) kink modulus leads to formation of towers of such states. 
	For the $ \mc{Z}_N $-symmetric twisted masses
	their number is by far less than $ N - 1 $ as was conjectured previously.
	We investigate  the quasiclassical limit and show that out of $ N $ possible towers  
	only two survive in the spectrum for odd $ N $, and a single tower for even $ N $.
	In the case of CP$^2$ theory the related CMS are discussed in detail.
%	The tower of the bound states, which is the heavier of the two, encounters multiple CMS and
%	disappears before reaching the strong coupling region.
%	The lighter tower decays on the primary CMS leaving only two states which continue through
%	into strong coupling.
	In these points we overlap and completely agree with the results of  Dorey and Petunin.
We also comment on 2D--4D correspondence.

\end{titlepage}

\newpage

%%%%%%%%%%%%%%%%%%%%%%%%%%%%%%%%%%%%%%%%%%%%%%%%%%%%%%%%%%%%%%%%%%%%%%%%%%%%%%%%%%
%                                                                                %
%                      T A B L E   O F   C O N T E N T S                         %
%                                                                                %
%%%%%%%%%%%%%%%%%%%%%%%%%%%%%%%%%%%%%%%%%%%%%%%%%%%%%%%%%%%%%%%%%%%%%%%%%%%%%%%%%%
\tableofcontents

\newpage

%%%%%%%%%%%%%%%%%%%%%%%%%%%%%%%%%%%%%%%%%%%%%%%%%%%%%%%%%%%%%%%%%%%%%%%%%%%%%%%%%%
%                                                                                %
%                          I N T R O D U C T I O N                               %
%                                                                                %
%%%%%%%%%%%%%%%%%%%%%%%%%%%%%%%%%%%%%%%%%%%%%%%%%%%%%%%%%%%%%%%%%%%%%%%%%%%%%%%%%%
\section{Introduction}
\setcounter{equation}{0}

	The observation of the exact nonperturbative coincidence between the BPS spectrum
	of the kinks in the two-dimensional CP$^{N-1}$ models with twisted masses and the BPS spectrum of the 
	monopole-dyons
	in the four-dimensional Seiberg--Witten solution 
\cite{SeibergWitten}
	dates back to 1998 
\cite{Dorey:1998yh}. On the four-dimensional side this coincidence is for a particular vacuum of ${\mathcal N}=2$
supersymmetric QCD with gauge group U$(N)$ in which a maximal number $r=N$ of quark flavors condense.
	This observation was understood and interpreted in clear-cut physical terms after the discovery of 
	the non-Abelian strings 
	\cite{1,2} in the $r=N$ Higgs vacuum of ${\mathcal N}=2$ Yang--Mills and the discovery of confined monopoles attached to them 
\cite{Shifman:2004dr,4} 
	(represented by kinks in the effective theory on the string world sheet). 
	It is crucial that the monopole-dyon mass  on the Coulomb branch of super QCD in the $r=N$ vacuum is given by the same formula  
	as the mass in the presence of  the  Fayet--Iliopoulos term.
	The underlying reason was explained in 
\cite{Shifman:2004dr,4}. 
	The above results  set the stage for the development of the 2D--4D correspondence in this particular problem.

	An immediate consequence of the spectral coincidence in two and four dimensions 
	(in the Bogomol'nyi--Prasad--Sommerfield 
\cite{BPS}, 
	BPS for short, sectors) 
	is the coincidence of the curves of marginal stability (CMS) or the  wall crossings.
	These curves, being established in two dimensions for kinks can be immediately elevated to four dimensions, for monopole-dyons. 
	Generally speaking, for non-Abelian strings in the SU$(N)$ gauge theories there are $N$ appropriate mass parameters.
	Correspondingly, there are $N$ twisted mass parameters in the two-dimensional CP$^{N-1}$ models (considered in the gauged formulation). 
	Each mass parameter is a complex number. 
	Therefore, in the general case one has to deal with highly multidimensional 
	 wall crossings.

	In \cite{Gorsky} it was pointed out that extremely beneficial for various physical applications 
	was a special choice 
\beq
\label{znmasses}
	m_k ~~=~~ m_0 \cdot e^{2 \pi i k / N}\,,\qquad\qquad k~=~0\,,~1\,,~ ...\,,~ N-1\,,
\eeq
	preserving a discrete $Z_N$ symmetry of the model. 
	Then there is only one adjustable mass parameter $m_0$,
	and the  wall crossings reduce to a number of CMS on the complex plane of $m_0$.
	We will refer to \eqref{znmasses} as the $Z_N$ symmetric masses.

	These CMS were studied several times in the past. For CP$^1$ the result was found in \cite{SVZw}.
	In this case the solution is clear and simple; no questions as to its validity arise.
	A more contrived case of 
	CP$^{N-1}$ (with $N\geq 3$) was addressed in \cite{5,Bolokhov:2011mp}. 
	A detailed study carried out in \cite{Bolokhov:2011mp} revealed
	the existence of a richer kink spectrum than was originally anticipated \cite{Dorey:1998yh}. 
	In particular, the existence of $ N-1 $ towers was argued replacing  a single tower discussed in \cite{Dorey:1998yh}. 
	However, shortly after, we realized
	that we had analyzed only necessary conditions for these towers to exist, leaving the sufficient conditions aside.
	In this paper we close this gap. 
	In the weak-coupling sector, we find that, although (based on the analyses of the central charges) $ N-1 $ towers could exist, 
	in fact, the sufficient conditions are met only for two towers for odd $N$ and a single tower for even $N$. 
	A generic form of the sufficient conditions were discussed long ago \cite{Dorey:1999zk}.
	The second tower for odd $N$ appears as a collection of bound states of the appropriate kinks with a quantum of
	an elementary excitation carrying no topological number.  In four-dimensional language this is a bound state of a monopole with (s)quark/gauge boson.

	The ``extra'' strong-coupling states identified previously do not survive when passing from strong coupling
	into the weak coupling.
	This is so for CP$^{N-1}$ with even $ N $ as there are no corresponding bound states.
	In addition, we showed that for $ N = 3 $ the only extra state decays in passing from strong 
	to weak coupling. 
	Two other strong-coupling states form the basis for the first tower. 
	The second tower appears only at weak coupling and decays in passing from weak to strong coupling.
	It is plausible to conclude that for generic $ N $ the ``extra'' states at strong coupling
	and the extra tower of bound states at weak coupling are independent.

	On the four-dimensional side the general theory of the wall crossing was worked out in \cite{koso}. 
	Its consequences for the monopole-dyons in the case of the $Z_N$ symmetric masses \eqref{znmasses}
	are studied in\,\footnote{We are grateful to the authors for providing us with a draft of their paper prior to its publication.}
	\cite{ndkp}. 
	Our task is to analyze the kink spectrum and the corresponding CMS in two dimensions, 
	taking account of the sufficient conditions mentioned above, 
	and compare the result with that in four dimensions, 
	thus  demonstrating the power of 2D--4D correspondence. We should emphasize that the
	CMS pattern for $N=3$ is much more contrived than that for $N=2$.
	
	Our strategy will be as follows. 
	We will focus on the first nontrivial case, namely CP$^2$ model,
	in which we will carry out a complete quasiclassical analysis of the kink (dyon) bound states   with elementary
	excitation quanta at large masses. 
	We then determine all relevant curves of marginal stability and 
	explain the survival of three Hori--Vafa \cite{MR1} states in the physical spectrum at small (or vanishing) twisted masses.
	Then we will generalize the lessons obtained in CP$^2$ to higher values of $N$.
	
	The paper is organized as follows.  
	Section  \ref{2D4D} describes the general setting and formulation of the problem and 
	outlines some facts known from the previous investigations. 
	In Section 3 we present an exhaustive analysis of the 
	Veneziano--Yankielowicz superpotential (which is exact in the given problem) --- 
	namely we study its analytical properties, find the Argyres--Douglas points and so on.	
	Section \ref{qclassics} presents a thorough analysis of the quasiclassical limit (large mass terms).
	This is the domain of weak coupling. 
	Here we explicitly built the kinks in the CP$^N$ model, and 
	identify the bound states of the kinks with the elementary quanta. 
	Section \ref{secspectrum} is devoted to the study of the BPS-sector spectrum and 
	the curves of the marginal stability in the CP$^N$ model with $Z_3$ symmetric twisted masses.

%%%%%%%%%%%%%%%%%%%%%%%%%%%%%%%%%%%%%%%%%%%%%%%%%%%%%%%%%%%%%%%%%%%%%%%%%%%%%%%%%%
%                                                                                %
%                           P R E L I M I N A R I E S                            %
%                                                                                %
%%%%%%%%%%%%%%%%%%%%%%%%%%%%%%%%%%%%%%%%%%%%%%%%%%%%%%%%%%%%%%%%%%%%%%%%%%%%%%%%%%
\section{2D\,--\,4D Correspondence and Preliminaries}
\setcounter{equation}{0}
\label{2D4D}

%%%%%%%%%%%%%%%%%%%%%%%%%%%%%%%%%%%%%%%%%%%%%%%%%%%%%%%%%%%%%%%%%%%%%%%%%%%%%%%%%%
%                  2 D   --  4 D   C O R R E S P O N D E N C E                   %
%%%%%%%%%%%%%%%%%%%%%%%%%%%%%%%%%%%%%%%%%%%%%%%%%%%%%%%%%%%%%%%%%%%%%%%%%%%%%%%%%%
\subsection{2D\,--\,4D correspondence }

	As was shown in \cite{Dorey:1998yh},
	the BPS spectrum of dyons (at the singular point on the Coulomb branch 
	in which  $ N $ quarks become massless) in the
	four-dimensional \ntwo supersymmetric QCD with gauge group U$(N)$ and  $ N_f\,=\,N $ fundamental flavors (quarks), identically
	coincides with the BPS spectrum in the two-dimensional twisted-mass deformed CP$(N-1)$ model.
	The reason for this coincidence was revealed 
	in \cite{Shifman:2004dr,4}. 
	Here we briefly review this coincidence.

	Non-Abelian strings \cite{1,2} were first found in \ntwo supersymmetric QCD with gauge group U$(N)$ and  $N_f=N$ quarks 
	with a Fayet--Iliopoulos term $\xi$, see \cite{Trev,Jrev,SYrev,Trev2} for a review.
	They were found in a particular vacuum where the maximal number of quarks $r=N$ condense (at non-zero $\xi$).
	In this vacuum the scalar quarks (squarks) develop condensate which results in  the spontaneous
	breaking of both the gauge U($N$) group and flavor (global) SU($N_f=N$) group, leaving unbroken a
	diagonal global SU$(N)_{C+F}$,
\beq
{\rm U}(N)_{\rm gauge}\times {\rm SU}(N)_{\rm flavor}
\to {\rm SU}(N)_{C+F}\,.
\label{c+f}
\eeq
	Thus, a color-flavor locking takes place in the vacuum.
	The presence of the global SU$(N)_{C+F}$ group is a key reason for the
	formation of non-Abelian strings whose main feature is
	the occurrence  of orientational zero modes associated with rotations of the flux
	inside the  SU$(N)_{C+F}$ group. 
	Dynamics of these orientational moduli are described by
	the effective two-dimensional   \ntwot supersymmetric CP$(N-1)$ model on
	the string world sheet \cite{1,2,Shifman:2004dr,4}.

	\ntwo supersymmetric QCD also has adjoint scalar fields which along with the gauge fields form the \ntwo vector supermultiplet. 
	These  adjoint scalars develop  VEVs as well, Higgsing the gauge U$(N)$ group
	down to its maximal Abelian subgroup if quark masses chosen to be different.
	This ensures existence of the conventional  't Hooft-Polyakov monopoles in the theory.
	The squark condensates then break the gauge U$(N)$ group completely, Higgsing all gauge bosons.
	Since  the gauge group is fully Higgsed the 't Hooft-Polyakov monopoles are {\em confined}.  
	In fact, in the   U$(N)$ gauge theories they are presented by junctions of two  different elementary non-Abelian strings.
	Since $N$ elementary non-Abelian strings correspond to $N$ vacua of the world-sheet theory,
	the   confined monopoles of the bulk theory are seen as kinks in the world-sheet 
	theory \cite{Tong,Shifman:2004dr,4}.

	Although the 't Hooft--Polyakov monopole on the Coulomb branch
	looks very different from the string junction of the theory in the Higgs regime,
	amazingly, their masses are the same 
	\cite{Shifman:2004dr,4}. 
	This is due to the fact that the mass of the BPS states (the string junction is a 1/4-BPS state) cannot depend on
	$\xi$ because $\xi$ is a nonholomorphic parameter. 
	Since the confined monopole emerges in the world-sheet theory as a kink, the Seiberg--Witten
	formula for its mass should coincide with the exact result for the kink
	mass in two-dimensional \ntwo twisted-mass deformed  CP$(N-1)$ model, which is 
	the world-sheet theory for the non-Abelian string in the bulk theory with $N_f=N$. 
	Thus, we arrive at the statement of coincidence of the BPS spectra in both theories.

%%%%%%%%%%%%%%%%%%%%%%%%%%%%%%%%%%%%%%%%%%%%%%%%%%%%%%%%%%%%%%%%%%%%%%%%%%%%%%%%%%
%                           P R E L I M I N A R I E S                            %
%%%%%%%%%%%%%%%%%%%%%%%%%%%%%%%%%%%%%%%%%%%%%%%%%%%%%%%%%%%%%%%%%%%%%%%%%%%%%%%%%%
\subsection{Preliminaries}
%\setcounter{equation}{0}
%\addcontentsline{toc}{subsection}{Prerequisites}
\label{prer}

	In our previous paper \cite{Bolokhov:2011mp} for the sake of defining the spectrum we introduced a function $ U_0(m_0) $,
\begin{align}
	U_0 (m_0) & ~~=~~ \left(\, e^{2 \pi i / N} ~-~ 1 \,\right) \cdot \mc{W}(\sigma_0)\,
	~~=~~
\nonumber
	\\
	&
	~~=~~ -\, \frac{1}{2\pi} \lgr e^{2\pi i / N} \,-\, 1 \rgr \biggl\{\, N \sqrt[N] { m_0^N \,+\, \Lambda^N }  
	\\
\nonumber
	&
	~~+~~  \sum_j\, m_j\, \ln \, \frac{ \sqrt[N] { m_0^N \,+\, \Lambda^N } \,-\, m_j } { \Lambda} \,\biggr\}\,,
\end{align}
	which we argued to be a single-valued quantity in a physical region of the mass parameter $ m_0 $.
	The $\mc{Z}_N$ invariance divides the complex plane of $ m_0 $ into $ N $ physically equivalent  sectors,
	each having the angle $2\pi/N$. Each sector covers the entire complex plane of
	$m_0^N$. 
	It follows, say, from the $\theta$ dependence that $m_0^N$ is the appropriate physical parameter.
	
	We demonstrated that the mirror-symmetry analysis requires $ N $ states to exist at strong coupling,
	with  the 
	central charges
\beq
\label{Usp}
	\mc{Z}  ~~=~~  U_0(m_0)  ~~+~~  i\,m_k\,.
\eeq
	We argued that continuations of these states to  weak coupling should be promoted to ``towers''
	corresponding to nonminimal values of the U(1) kink modulus.
	The question remained open at that time was as follows: ``whether or not these states represent
	actual bound states in the physical spectrum, and --- if yes --- what are their dynamical components?"

	Before turning to these questions,  let us first show what the moduli space of the parameter
	$ m_0 $ looks like in the CP$^{N-1}$ models, and, in particular, CP$^2$ (i.e. $ N \,=\, 3 $).

%%%%%%%%%%%%%%%%%%%%%%%%%%%%%%%%%%%%%%%%%%%%%%%%%%%%%%%%%%%%%%%%%%%%%%%%%%%%%%%%%%
%                                                                                %
%                         E X A C T   A N A L Y S I S                            %
%                                                                                %
%%%%%%%%%%%%%%%%%%%%%%%%%%%%%%%%%%%%%%%%%%%%%%%%%%%%%%%%%%%%%%%%%%%%%%%%%%%%%%%%%%
\section{Exact Analysis}
\setcounter{equation}{0}
\label{exact}

	In the \ntwo supersymmetric CP$^{N-1}$ theory one has the exact superpotential which
	describes the entire BPS  spectrum nonperturbatively. It is given by \cite{W93,AdDVecSal,ChVa,HaHo}
	
\beq
	\W (\sigma) ~~=~~ \frac{1}{2\pi}\,
	\sum_j\,(\sigma - m_j)\, \lgr \ln \frac{\sigma_p \;-\; m_j}{\Lambda} ~-~ 1  \rgr\,.
\eeq

	In this normalization the vacuum values of the superpotential take the form
\beq
\label{Wvac}
	\W ( \sigma_p ) ~~=~~ 
		-\, \frac{1}{2\pi}\,  
                \Bigl\{\, N\, \sigma_p ~+~ \sum_j\, m_j\, \ln \,\frac{\sigma_p - m_j}{\Lambda} \,\Bigr\}\,.
\eeq
	Here and in the remainder of this paper we put the (nonperturbative) dynamical scale parameter $ \Lambda $ to one.
	In Eq.~\eqref{Wvac}, $ \sigma_p $ is the position of the $ p $-th vacuum.
	In our case of the $ \mc{Z}_N $-symmetric twisted masses, the vacua sit at
\beq
\label{sig}
	\sigma_p^N  ~~=~~ 1 ~~+~~ m_0^N \,.
\eeq
	All these quantities become functions of a single parameter $ m_0 $.
	The spectrum, {\it i.e.} the central charges of both perturbative and nonperturbative states,
	is given by the differences of the vacuum values of the superpotential.
	The masses of the elementary BPS kinks will be given by the differences of the superpotential in
	two neighbouring vacua.
	Because of the $ \mc{Z}_N $ symmetry of the problem one can always choose the latter to be 
	the 0$^\text{th}$ and the 1$^\text{st}$ vacua,  i.e. $ p ~=~ 0,\; 1 $.
	The masses of the elementary states ({\it i.e.} states with no topological charge) are obtained 
	as differences of the superpotential in the same vacua, but with the mass parameter $ m_0 $ 
	sitting on different branches of the relevant logarithms.

	The problem with these expressions, as emphasized in \cite{HaHo,SYtorkink,Bolokhov:2011mp}, is in their 
	multiva\-luedness.
	This appears both in the logarithms in Eq.~\eqref{Wvac} as well as in the $ N $-th root in Eq.~\eqref{sig}.
	The correct handling of the superpotential involves analysis of the whole complex manifold
	of the mass parameter.
	In the case of a general $ N $ such a manifold can have a rather complicated structure.
	We make a few general statements regarding the superpotential in the CP$^{N-1}$ model, and then pass
	to building the entire complex manifold in the case of CP$^2$.

%%%%%%%%%%%%%%%%%%%%%%%%%%%%%%%%%%%%%%%%%%%%%%%%%%%%%%%%%%%%%%%%%%%%%%%%%%%%%%%%%%
%              S M A R T   A N D   S I L L Y   L O G A R I T H M S               %
%%%%%%%%%%%%%%%%%%%%%%%%%%%%%%%%%%%%%%%%%%%%%%%%%%%%%%%%%%%%%%%%%%%%%%%%%%%%%%%%%%
\subsection{Prerequisites: ``smart" and ``silly" logarithms}

	As was pointed out, the expression for $ \mc{W}(\sigma_0) $ is quite ambiguous.
	Even more so is the expression for the difference
\beq
\label{diff}
	\mc{W}(\sigma_1) ~~-~~ \mc{W}(\sigma_0)\,,
\eeq
	which describes the physical masses of the BPS states.
	We need to correctly pin-point what is understood when we speak of logarithms 
	and $ N $-th roots in our expressions.
	We can picture two schemes how one can do this. 

	We repeat that the entire spectrum is contained in the multivalued expression \eqref{diff}.
	The mass parameter $ m_0 $ can travel through different branches of the logarithms, 
	and, in the first approach, the latter change as continuous functions.
	Indeed, the logarithms defined on multiple branches do not have any discontinuities;
	this is the reason why the branches are introduced in the first place. 
	These are the ``smart'' logarithms, which are supposed to give us all physical states. 
	They are even slightly ``smarter'' than ordinary multibranched logarithms, since they also should tell us
	which states  obtained in this way are physically stable and which are not. 

	Such quantities are difficult to deal with. 
	The other approach to handle the multivaluedness is, as usual, to introduce branch cuts on a single complex plane. 
	We adopt the following definition of the function $ \log~x $:
\beq
	\log~x ~~\in~~ \mc{R}\,, \qquad\qquad \text{when~~} x ~>~ 0\,.
\eeq
	We draw a branch cut of this logarithm to extend from the origin to the negative infinity,
\beq
	\text{branch cut:}\qquad -\infty ~~<~~ x ~~<~~ 0\,,
\eeq
	and put the argument of complex variable to be in the usual domain,
\beq
	-\,\pi ~~<~~  \text{Im}~\log x  ~~\leq~~ \pi\,.
\eeq
	This function is defined on a single complex plane and it does not know about various branches.
	For this reason such quantities can be called ``silly'' logarithms.
	Whenever $ x $ passes to a different branch, one needs to add a $ 2 \pi i $ explicitly.
	These logarithms have a discontinuity across the branch cut.

	To an extent a similar statement applies to the definition of $ \sigma_p $.
	We define the physical branch of the $ N $-th power root by insisting that
	at large values of the twisted masses, the vacua $ \sigma_p $ take on their (quasi)classical values,
\beq
\label{sigclass}
	\sigma_p  ~~\approx~~ m_p\,,  \qquad\qquad\qquad |m_p| ~\gg~ 1\,.
\eeq
	Stepping out of that branch leads to the same spectrum obtained with a permutation of the vacua.
	In this sense, all $ N $ branches of the $ N $-th power root are ``physical'', but they all have the
	same spectra, just the vacua re-named.
	Although in principle, the analysis of the whole complex manifold should include 
	all physically equivalent branches of the root, we do not do this, 
	limiting ourselves to a single physical branch.
	We denote the $ N $-root as a function on a single complex plane,  
\beq
	\sqrt[N]{x} ~~>~~ 0\,, \qquad\qquad \text{when~~} x ~>~ 0\,,
\eeq
	and define
\beq
	\sigma_p ~~=~~ e^{2 \pi i p / N} \cdot \sigma_0 ~~=~~ e^{2 \pi i p / N} \cdot \sqrt[N]{ 1 ~+~ m_0^N }\,.
\eeq
	The branch cuts of this root in the $ m_0 $ plane should be drawn in such a way as to preserve
	the classical relation \eqref{sigclass}.

	We reiterate that the physical values of the central charge 
\beq
\label{ccharge}
	\mc{Z} ~~=~~ \W(\sigma_p) ~~-~~ \W(\sigma_q)
\eeq
	are given by the ``smart'' logarithms in the functions $ \W(\sigma) $.
	In the remainder  of the paper, however, we will define the superpotential $ \W(\sigma_p) $
\beq
\label{Wv}
	\W ( \sigma_p ) ~~=~~ 
		-\, \frac{1}{2\pi}\,  
                \Bigl\{\, N\, \sigma_p ~+~ \sum_j\, m_j\, \log\, \big( \sigma_p - m_j \big) \,\Bigr\}
\eeq
	as a function of ``silly" logarithms. 
	For each moduli manifold, we therefore will need to introduce a set of branch cuts and 
	specify how the logarithms should jump in order to account for passing onto a different branch.
	Each such jump, proportional to the mass $ m_j $, should imitate the ``smart'' logarithm,
	which is continuous.
	The jump is introduced precisely to compensate for the discontinuity of the ``silly'' logarithms.
	As defined, the function \eqref{Wv} does not know which branch we are currently sitting on.
	Therefore, we will need to also specify a reference point from which we start accounting.
	All the rest places of the manifold are reached by passing from this reference point along a contour.
	While passing along the contour, we pick up all the jumps that occur on the ``edges'' of 
	branches, while, again, \eqref{Wv} does not know that we are in fact jumping from
	branch to branch.

%%%%%%%%%%%%%%%%%%%%%%%%%%%%%%%%%%%%%%%%%%%%%%%%%%%%%%%%%%%%%%%%%%%%%%%%%%%%%%%%%%
%          C O M P L E X   M A N I F O L D   O F   M   I N   C P ^ { N - 1 }     %
%%%%%%%%%%%%%%%%%%%%%%%%%%%%%%%%%%%%%%%%%%%%%%%%%%%%%%%%%%%%%%%%%%%%%%%%%%%%%%%%%%
\subsection[Complex manifold of $ m $ in CP$^{N-1}$]
	{Complex manifold of \boldmath{$ m $} in CP\boldmath{$^{N-1}$}}

	The  complex manifold for a general CP$^{N-1}$ case will be rather complicated and include
	multiple branch cuts of the logarithms.
	We do not perform a full analysis of all such manifolds, which will also depend on 
	the parity of $ N $.
	But we still make a few general statements regarding these manifolds and the behaviour of function $ \W(\sigma_p) $.
	For simplicity, we denote it as
\beq
	\W_p ~~\stackrel{\text{def}}{=} ~~ \W(\sigma_p)\,,
\eeq
	and always remember that it is a function of $ m_0 $.

\begin{figure}
\begin{center}
\epsfxsize=8.0cm
\epsfbox{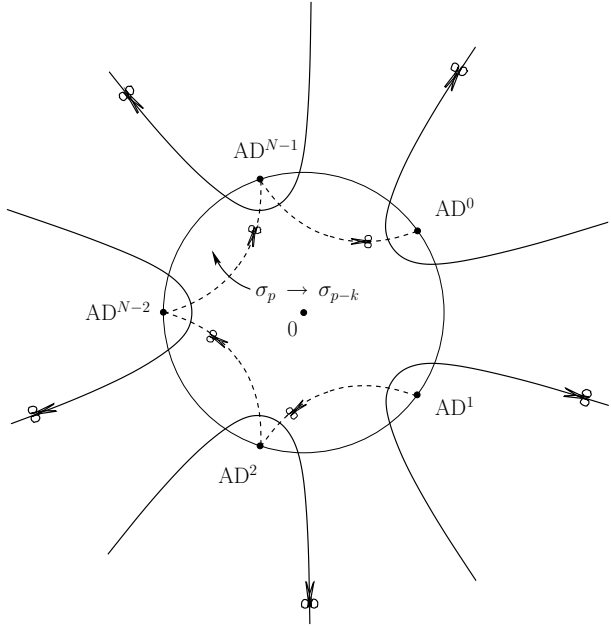}
\caption{General appearance of a complex moduli space of $ m $ in the case of CP$^{N-1}$}
\label{fcpn}
\end{center}
\end{figure}
	Figure~\ref{fcpn} shows a schematic appearance of a moduli space in CP$^{N-1}$.
	The complex plane is diced by various branch cuts of the logarithms.
	Solid lines sketch the logarithmic branch cuts, dashed lines show the branch cuts of the $N$-root in $\sigma$.
	We immediately comment that the logarithm branch cuts are non-compact, 
	{\it i.e.} they extend from infinity to infinity.
	This is related to the fact that the logarithms depend on combinations $ \sigma_p \,-\, m_j $,
	rather than just on masses $ m_j $ themselves.
	Branch cuts appear in the places where these differences become real and negative.
	As such differences never vanish for finite $ m_0 $, the branch cuts may terminate only at infinity. 
	In general, each branch cut is shared by multiple logarithms.

	There are $ N $ Argyres-Douglas (AD) points \cite{AD} which are located at positions
\beq
	m_0^\text{AD} \,: \qquad\quad \left( m^\text{AD} \right)^N ~~=~~ -1 \,.
\eeq
	The reason for the particular numbering of the AD points shown in the figure will become clear in just a moment.
	The points are enclosed by ``cups'' of hyperbolic-shaped logarithmic branch cuts.
	There are also $ N - 1 $ branch cuts of the $ N $-root coming from the definition of $ \sigma_p $ \eqref{sig}.
	They can be chosen so as to connect the AD points, as shown by the dashed lines.
	As one crosses such a branch cut, the phase of each $ \sigma_p $ experiences a jump, 
	coherent to classical relation \eqref{sigclass}. 
	If one starts at the origin and crosses the branch cut connecting AD$^{N-1}$ and AD$^0$, 
	the vacua shift as $ \sigma_p \,\to\, \sigma_{p-1} $.
	If one instead crosses the branch cut connecting AD$^{N-2}$ and AD$^{N-1}$,
	the shift will be $ \sigma_p \,\to\, \sigma_{p-2} $.
	Following further counter-clockwise, if one crosses the $ k $-th branch cut 
	that connects AD$^{N-k}$ to AD$^{N-k+1}$ when passing from the inside outside, the vacua shift by $ k $ turns
\beq
\label{sigshift}
	\sigma_p ~~\to~~ \sigma_{p-k} \,.
\eeq
	In accord with this, a branch cut of logarithm $ \log\;( \sigma_p \,-\, m_j ) $ turns into 
	a branch cut of $ \log\;( \sigma_{p \pm k} \,-\, m_j ) $ after it crosses the $ k $-th branch cut of $ \sigma $. 
	The plus or minus sign will of course depend on the direction of this crossing. 
	The branch cut of $ \log\;( \sigma_p \,-\, m_j ) $ itself comes off the physical plane.

\begin{figure}
\begin{center}
\epsfxsize=8.0cm
\epsfbox{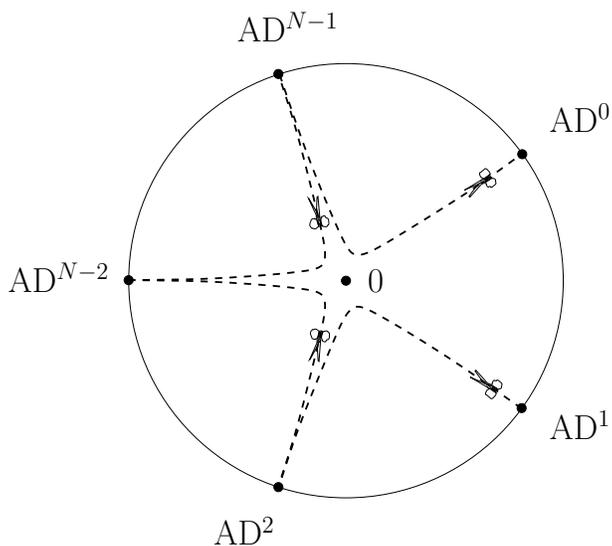}
\caption{Stretched branch cuts of $ \sigma $ in CP$^{N-1}$.}
\label{fsigcpn}
\end{center}
\end{figure}

	For what follows, it is convenient to deform the branch cuts of $ \sigma $ so that they almost touch the origin.
	Each cut will start from an AD point, come infinitesimally close to the origin in a straight line, 
	then turn around and follow to the other AD point also in a straight line, this way making a letter ``v''.
	The system of the $ \sigma $ branch cuts will then form a star with the center at the origin, as sketched in Fig.~\ref{fsigcpn}.
	The convenience of such a configuration of cuts now shows up in a simple relation,
\beq
\label{mjump}
	\text{as}~~ m_0 ~~\to~~ e^{2\pi i / N}\, m_0\,,  \qquad\qquad \sigma_p ~~\to~~ e^{2\pi i / N} \sigma_p\,.
\eeq
	That is, as $ m_0 $ jumps from one ``sector'' to another, the phase of $ \sigma_p $ jumps accordingly.
	That this is so at the weak coupling $ |m_0| \,\gg\, 1 $ is not a surprise, but a consequence of Eq.~\eqref{sigclass}.
	In the region of small $ m_0 $, however, this did not have to be so, and we have had to arrange the branch cuts 
	in a special way so that relation \eqref{mjump} is retained.

	Besides introducing the branch cuts, we also need to specify a reference point from which we start counting
	the central charge.
	We notice that, any AD point would do in particular.
	Say we pick AD$^{(p)}$.
	Then we automatically define the central charge of an elementary kink to vanish at AD$^{(p)}$,  just because
	all $ \sigma $'s are equal at an AD point.
	The value of the central charge in other places, in particular at other AD points, is obtained by following
	a path connecting the reference point and the destination point, and accounting for the branch cuts
	along the way. 
	The central charge of a kink will not vanish at other AD points. 

	Since our problem has a $ Z_N $ symmetry, the complex plane of $ m_0 $ is split into $ N $ equivalent physical
	regions.
	Each physical region contains exactly one AD point.
	It does not matter where to start measuring the regions from.
	Because of the different branch cuts, the regions might seem non-equivalent to each other.
	The equivalence is of course reflected in the coincidence of the physical spectra in all these regions.

	The description of the complex space given above is enough for us to introduce two relations which have
	direct connection to $ \mc{Z}_N $ symmetry of the theory.
	We notice that the superpotential
\beq
	\W_p ~~=~~ 
		-\, \frac{1}{2\pi}\,  
                \Bigl\{\, N\, \sigma_p ~+~ \sum_j\, m_j\, \ln\, \big( \sigma_p - m_j \big) \,\Bigr\}
\eeq
	seemingly has a $ \mc{Z}_N $ symmetry of its vacuum values. 
	Is it true indeed that the vacuum values of $ \W $ sit on a circle?
	The answer is no. 
	Instead, the differences of the vacuum values do,
\beq
\label{Wdiff}
	\W_{p+2}  ~~-~~  \W_{p+1}  ~~=~~ 
        e^{2 \pi i / N}\, \Bigl(\,  \W_{p+1}  ~~-~~ \W_{p}   \,\Bigr) \,.
\eeq
	Similar relations for the vacuum values themselves depend on the location in the moduli space.
	For us, immediate vicinities of the AD points are important reference locations.
	The branch cuts of the logarithms form ``cups'' around each of the AD points.
	Inside each of the cup, the following relations can be inferred,
\beq
\label{shift}
	\W_{p+1} ~~=~~ e^{2 \pi i / N} \cdot \W_p ~~+~~ i\, m_s\,,
\eeq
	and
\beq
\label{shiftm}
	\W_p\, (e^{2 \pi i / N} \cdot m_0) ~~=~~ e^{2 \pi i / N} \cdot \W_p\, (m_0) ~~+~~ i\, m_s\,.
\eeq
	Here $ s = 0 $ for AD$^{(0)}$, $ s = 1 $ for AD$^{(1)}$ {\it etc}, gives the numeration
	to AD points as shown in Fig.~\ref{fcpn}.

	As an application, Eq.~\eqref{shift} immediately gives us the values of $ \W_p $ at AD points.
	Indeed, at any such point $ \sigma_p \,=\, 0 $ and so all $ \W_p $ are equal.
	Therefore,
\beq
	\left(\, e^{2 \pi i / N} \;-\; 1 \,\right)\, \W_p\, \biggr|_\text{AD$^{(s)}$} ~~=~~ -\, i\, m_s\,.
\eeq
	The right hand side is different for all AD points, seemingly. 
	Upon a closer look, the value of $ m_s $ at the $ s $-th AD point is independent of $ s $ and equals
	$ |m_0|\,e^{i \pi / N} $.

%%%%%%%%%%%%%%%%%%%%%%%%%%%%%%%%%%%%%%%%%%%%%%%%%%%%%%%%%%%%%%%%%%%%%%%%%%%%%%%%%%
%                T H E   M O D U L I   S P A C E   O F   C P ^ 2                 %
%%%%%%%%%%%%%%%%%%%%%%%%%%%%%%%%%%%%%%%%%%%%%%%%%%%%%%%%%%%%%%%%%%%%%%%%%%%%%%%%%%
\subsection[{The moduli space of CP$^2$}]
	{The moduli space of CP\boldmath{$^2$}}

\begin{figure}
\begin{center}
\epsfxsize=7.5cm
\epsfbox{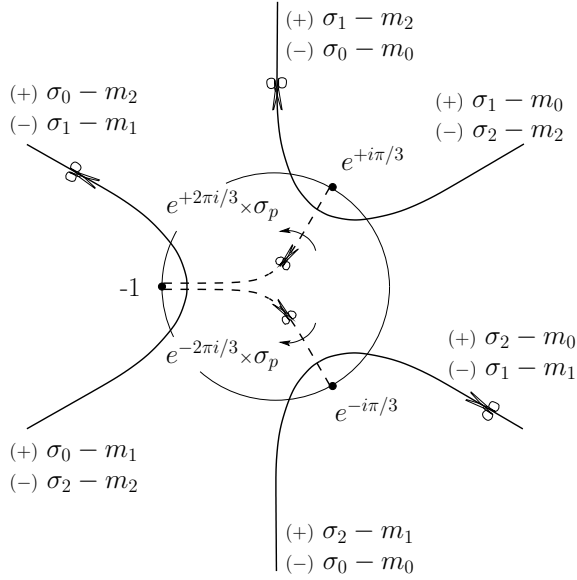}
\caption{Moduli space of CP$^2$}
\label{fcp2}
\end{center}
\end{figure}

	The moduli space in the case of CP$^2$ is shown in Fig.~\ref{fcp2}. 
	It has hyperbolic branch cuts of the logarithms, determined by the equation
\beq
	\text{Re}~m^3 ~~=~~ -\,\frac{1}{2}\,.
\eeq
	The branch cuts belonging to the physical plane are marked with subscripts which
	identify them to their respective logarithms.
	The $ {\scriptstyle (+)} $ or $ {\scriptstyle (-)} $ signs indicate 
	the change ($ + 2 \pi i $ or $ - 2 \pi i $) of a logarithm 
	when its branch cut is crossed counter-clockwise.
	Note that each line is a branch cut of two different logarithms. 
	The branch cuts of $ \sigma $, as discussed previously, can be chosen to connect the AD points.
	As one crosses a branch cut of $ \sigma $ from right to left (indicated by arrows in Fig.~\ref{fcp2}), 
	the phase of each $ \sigma_p $ shifts by $ e^{+ 2 \pi i / 3} $ on the upper cut, 
	and by $ e^{- 2 \pi i / 3} $ on the lower cut.
	Somewhat counter-intuitively, 
	this shifts $ \sigma_p \;\to\; \sigma_{p-1} $ and $ \sigma_p \;\to\; \sigma_{p+1} $,
	see Eq.~\eqref{sigshift}.
	Consequently, as logarithm lines cross the $ \sigma $ branch cuts, they turn into branch
	cuts of other logarithms, in accord with Eq.~\eqref{sigshift}.

	As discussed earlier, the logarithmic cuts form cups around the AD points. 
	Inside these cups, the following relations hold for the vacuum values of the superpotential,
\beq
	\W_{p+1} ~~=~~ e^{2 \pi i / 3} \cdot \W_p ~~+~~ i\, m_s\,, \qquad\quad s~=~0,~1,~2\,.
\eeq
	and
\beq
	\W_p\, (e^{2 \pi i / 3} \cdot m_0) ~~=~~ e^{2 \pi i / 3} \cdot \W_p\, (m_0) ~~+~~ i\, m_s
	\qquad\quad s~=~0,~1,~2\,.
\eeq
	Everywhere {\it outside} these cups, the superpotential values {\it are} sitting on the circle,
\beq
	\W_{p+1} ~~=~~ e^{2 \pi i / 3} \cdot \W_p
	\qquad\qquad \text{(outside the cups)}.
\eeq
	In particular, this is true on the real positive axis.
	This circumstance is very helpful in identifying the states seen as kinks in the mirror description.

	At this point we do not have yet an exact prescription how to choose a reference point.
	As we mentioned earlier, it is convenient to choose an AD point as reference.
	In fact all $ N $ AD points are equivalent and we can pick any of them.
	The ``star'' of the cubic-root branch cuts however gives us 
	an area $  -\pi ~<~ \text{Arg}\; m_0 ~<~ \pi $ free of cuts,
	which suggests us picking AD$^{(0)}$ or AD$^{(1)}$.
	To consider other AD points as reference points, 
	one would need to rotate the ``star'' of $\sigma$-branch cuts so that a corresponding region becomes
	free of cuts.
	This is not difficult, but not necessary either.
	So let us pick the reference point to be AD$^{(0)}$, 
\beq
	m_\text{ref} ~~=~~ m_\text{AD}^{(0)} ~~=~~ e^{i \pi / 3}\,.
\eeq
	We will justify this choice later.
	A state with the topological charge
\beq
	\vec{T} ~~=~~ (\, -1\,,~~ 1\,,~ 0 \,,~~ ...\,,~~ 0 \,)
\eeq
	becomes massless at this point.
	We shall denote this state with letter $ \mc{M} $ to symbolize its resemblance with 
	the four-dimensional monopole. 
	Such an interpretation will also be justified later in the paper.
	The central charge of this state is given by
\beq
\label{ZM}
	\mc{Z_M} ~~=~~ \W_1 ~~-~~ \W_0
	\qquad\qquad\qquad
	\text{(\,near AD$^{(0)}$\,)\,.}
\eeq
	This relation holds in the vicinity of AD$^{(0)}$.
	The central charge of this state on the real positive line is given by
\beq
\label{ZMR}
	\mc{Z_M} ~~=~~ \W_1 ~~-~~ \W_0 ~~+~~ i\, m_0
	\qquad\qquad\qquad
	\text{(\,on real axis\,)\,.}
\eeq
	Although seemingly disconnected, the two functions \eqref{ZM} and \eqref{ZMR} 
	continuously turn into each other.

	There are also dyonic states which are known to exist quasiclassically,
\beq
	\vec{T} ~~=~~ (\, -1\,,~~ 1\,,~ 0 \,,~~ ...\,,~~ 0 \,)\,, \qquad
	\vec{S} ~~=~~ (\, -n\,,~~ n\,,~ 0 \,,~~ ...\,,~~ 0 \,)\,, 
\eeq
	where $ n $ is an integer, and $ \vec{S} $ denotes the set of U(1) charges.
	The central charges of such states are given by 
\beq
	\mc{Z}_{\mc{D}^{(n)}} ~~=~~ \W_1 ~~-~~ \W_0\, ~~+~~ i\, n\, ( m_1 \,-\, m_0 ) \,.
\eeq
	This equation is valid near AD$^{(0)}$.
	We denote these dyonic states by $ \mc{D}^{(n)} $.

	We remind again that there are three physically equivalent regions, each one containing an AD point.
	The monodromies of the moduli space work in such a way that as one passes from 
	one physical region to the next one counter-clockwise, the difference 
$ \W_1 ~-~ \W_0 $
	gains one unit of the U(1) charge,
\beq
	\W_1 ~~-~~ \W_0  ~~~\to~~~ \W_1 ~~-~~ \W_0 ~~+~~ i\,( m_1 \,-\, m_0 ) \,.
\eeq
	Since the $ \M $ state becomes massless at AD$^{(0)}$, the state $ \D^{(1)} $ will become
	massless at AD$^{(1)}$, and the state $ \D^{(-1)} $ at AD$^{(2)}$.
	The paths connecting the initial point AD$^{(0)}$ and the destination points AD$^{(1)}$ and AD$^{(2)}$
	lie outside the unit circle region, and 
	extend clockwise and counter-clockwise correspondingly.
	If we further extend these paths through their AD points and into the circle towards the origin,
	the masses of the states will become
\beq
	\W_1 ~~-~~ \W_0 ~~+~~ i\, m_{1,2}\,.
\eeq
	Now if take the $ \M $ state and push it through the point AD$^{(0)}$ towards the origin,
	its mass will become $ \W_1 ~-~ \W_0 ~+~ i\,m_0 $, again, because of the branch cuts.
	All three states therefore can be written as
\beq
	\W_1 ~~-~~ \W_0 ~~+~~ i\, m_k\,,\qquad\qquad k~=~ 0,~1,~2\,.
\eeq
	Since in the area near the origin $ \W_1 ~=~ e^{2 \pi i / 3}\, \W_0 $, we obtain all three states 
	predicted by the mirror theory,
\beq
	(e^{2 \pi i /3} ~-~ 1)\, \W_0 ~~+~~ i\, m_k\,,\qquad\qquad k~=~ 0,~1,~2\,.
\eeq

	To clarify, these states are not the original dyonic states $ \D^{(n)} $ with $ n ~=~ -1 $,
	$ 0 $ and $ 1 $.
	They would be such if we had followed one and the same path for all three states from AD$^{(0)}$ to
	the origin, while we followed different paths.
	The procedure that we have done shows how the monodromies can be used to construct the spectrum
	seen in the mirror representation.

	We have prepared a mathematical ``background'' describing the moduli space and 
	correctly introduced the superpotential functions.
	We have also written out the three states existing at strong coupling as predicted by mirror symmetry.
	
	We now can analyze the CP$^{N-1}$ and, in particular, CP$^2$ problems quasiclassically, and 
	connect quasi-classical relations to the large-$ m $ expansion of exact functions $ \W(\sigma) $.
	This will enable us to establish the quantum numbers of both the weak- and strong-coupling states.
	We also will be able to build the spectrum of the bound states at weak coupling.

%%%%%%%%%%%%%%%%%%%%%%%%%%%%%%%%%%%%%%%%%%%%%%%%%%%%%%%%%%%%%%%%%%%%%%%%%%%%%%%%%%
%                                                                                %
%                          S E M I C L A S S I C S                               %
%                                                                                %
%%%%%%%%%%%%%%%%%%%%%%%%%%%%%%%%%%%%%%%%%%%%%%%%%%%%%%%%%%%%%%%%%%%%%%%%%%%%%%%%%%
%\newpage
\section{Quasiclassical Limit}
\label{qclassics}
\setcounter{equation}{0}

	Strong-coupling analysis by means of mirror symmetry predicts that in the neighbourhood of the origin
	there are $ N $ states.
	Can they be seen quasiclassically?
	The final answer will be given in Section~\ref{secspectrum}.
	It turns out that not all strong-coupling states exist in quasiclassics, while new, bound states appear.
	In this section we review the problem of finding the corresponding states at weak coupling, and 
	reconfirm the result of 
\cite{Dorey:1998yh}.
	In the case of $ \mc{Z}_N $ symmetric masses the result proves the existence of one bound state of a kink
	and an elementary quantum $ \psi_k $ with $ k ~=~ (N+1)/2 $ in the case when $ N $ is odd.
	Quantization of the U(1) coordinate of the kink then raises a tower of bound states of dyonic kinks
	and the elementary quantum.
	There are no bound states at weak coupling when $ N $ is even.

%%%%%%%%%%%%%%%%%%%%%%%%%%%%%%%%%%%%%%%%%%%%%%%%%%%%%%%%%%%%%%%%%%%%%%%%%%%%%%%%%%
%                         C E N T R A L  C H A R G E                             %
%%%%%%%%%%%%%%%%%%%%%%%%%%%%%%%%%%%%%%%%%%%%%%%%%%%%%%%%%%%%%%%%%%%%%%%%%%%%%%%%%%
\subsection{The central charge}

	The classical expression for the central charge has two contributions \cite{SVZw}:
	the Noether and the topological terms,
\beq
        \mc{Z} ~~=~~ i\, M_a\, q^a  ~~+~~ \int\, dz\, \p_z\, O \,, \qquad\qquad
	a ~=~ 1,\,...\, N-1\,.
	\label{21}
\eeq
	where $ M_a $ are the twisted masses (in the geometric formulation),
\beq
	M_a  ~~=~~ m_a ~-~ m_0\,,
\eeq
	$m^a$ ($a \,=\, 1,~2,~ ...,~N $) are the masses in the gauge formulation, 
	and the operator $O$ consists of two parts --- canonical and anomalous,
\begin{align}
	O & ~~=~~ 
	O_{\rm canon} ~~+~~ O_{\rm anom}\,,
\\[2mm]
	O_{\rm canon } 
\label{23}
	& ~~=~~ 
	\sum_{a= 1}^{N-1}\, M_a\, D^a\,,
\\[2mm]
	O_{\rm anom} 
\label{24}
	& ~~=~~
	-\, \frac{N\, g_0^2}{4\pi} 
	\lgr
	\sum_{a=1}^{N-1}\, M_a\, D^a ~+~ g_{i\bar{j}} \,\bpsi{}^\bj\, \frac{1-\gamma_5}{2}\,\psi^i
	\rgr.
\end{align}
	Moreover, the Noether charges $ q^a $ can be obtained from $N-1$ U(1) currents $ J_\mu^a $
	defined as\,\footnote{There is a typo in the definition of these currents in \cite{SVZw}.}
\begin{align}
	J_{RL}^a  
	& ~~=~~
	 g_{i\bj}\; \bphi^\bj\, (T^a)^{i\bj}\, i\overleftrightarrow{\p}{}_{\scriptscriptstyle \!\!\!\!RL}\, \phi^i   
\nonumber
	\\[2mm]
        & 
	~~+~~
        \frac{1}{2}\, g_{i\bj}\; \bpsi{}_{\scriptscriptstyle LR}^\bmm 
        	\lgr  ( T^a )_\bmm^{\ \bp}\, \delta_\bp^{\ \bj} ~+~ 
		\bphi{}^\br\, (T^a)_\br^{\ \bkk}\, \Gamma^{\;\bj}_{\bkk\bmm} \rgr   \psi_{\scriptscriptstyle LR}^i 
\nonumber 
	\\[2mm]
        & 
	~~+~~
        \frac{1}{2}\, g_{i\bj}\; \bpsi{}_{\scriptscriptstyle LR}^\bj
               \lgr  \delta^i_{\ p}\, (T^a)^p_{\ m} ~+~ 
	             \Gamma_{mk}^{\;i}\, (T^a)^k_{\ r}\, \phi^r \rgr   \psi_{\scriptscriptstyle LR}^m
\label{u1cur}
\end{align}
	in the geometric representation, and
\begin{align}
\notag
       J_{RL}^a  & ~~=~~ i\, \ov{n}{}_a \overleftrightarrow{\p}_{\scriptscriptstyle RL} n^a 
                   ~~-~~ |n^a|^2 \cdot i\, ( \ov{n} \overleftrightarrow{\p} n ) \\
                 & ~~+~~ \qquad 
                         \bxi{}_{\scriptscriptstyle LR}^a\, \xi_{\scriptscriptstyle LR}^a  ~~-~~
                         |n^a|^2 \cdot ( \bxi{}_{\scriptscriptstyle LR}\, \xi_{\scriptscriptstyle LR} )
\end{align}
	in the gauged formulation. 
	Here
 \beq
 	\left(T^a\right)^i_k = \delta_a^i\delta^a_k \,,\quad (\mbox{no summation over $a$!})
 \eeq
	and a similar expression for the overbarred indices.
	Finally,
$ D^a $ 
	are the Killing potentials,
\beq
	D^a  ~~=~~ r_0\, \frac{ \bphi\,\, T^a\, \phi } 
	                    {  1  ~+~  |\phi|^2  }
	     ~~=~~ r_0\, \frac{ \bphi^a\, \phi^a   }
	                    {  1  ~+~  |\phi|^2  }\,.
	                    \label{29}
\eeq
	The generators $ T^a $ always pick up the $ a $-th component.
	In this expression, 
\beq
	r _0  ~~=~~ \frac{ 2 }{ g_0^2}
\eeq 
	is a popular alternative notation for the sigma model coupling.
	
\begin{figure}
\begin{center}
\epsfxsize=8.0cm
\epsfbox{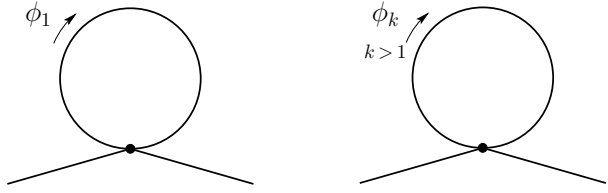}
\caption{Tadpoles contributing to the topological term $ O $.}
\label{ftad}
\end{center}
\end{figure}

	Note that Eq.~\eqref{29} contains the bare coupling. 
	It is clear that the one-loop correction must (and will) convert the bare coupling  into the renormalized coupling. 
	The anomalous part $O_{\rm anom}$ is obtained at one loop.
	Therefore, in the one-loop approximation for the central charge 
	it is sufficient to treat $O_{\rm anom}$ in the lowest order. Moreover, the bifermion term 
	in $O_{\rm anom}$ plays a role only in the two-loop approximation. 
	As a result, to calculate the central charge at one loop it is sufficient to analyze 
	the one-loop correction to $O_{\rm canon}$. 
	The latter is determined by the tadpole graphs in Fig~\ref{ftad}. 
	As usual, the simplest way to perform the calculation is to use the background field method.
	The part of the central charge under consideration is determined by the value of the fields at the spatial infinities. 
	In the CP$^2$ model to be considered below there are three vacua and three possible ways
	of interpolation between them. All kinks are equivalent. 
	We will be looking for the semi-classical expression for the central charge in the presence
	of the soliton interpolating between vacua ({\sc \small 0}) and ({\sc \small 1})
\beq
	\phi^1(z)  \,~~=~~\, e^{|M^1| z}\,, \qquad\qquad  \phi^2(z) \,~~=~~\, \phi^3(z) \,~~=~~ \,~...~\, ~~=~~\, 0\,.
\eeq
	That this is the right kink can be seen in the gauged formulation,
\begin{align}
\notag
	n^0  & ~~=~~ \frac {             1              }
 	                   {\sqrt{ 1 ~+~ e^{2 |M^1| z} }}\,, \\[3mm]
\notag
	n^1  & ~~=~~ \frac {         e^{|M^1| z}        }
 	                   {\sqrt{ 1 ~+~ e^{2 |M^1| z} }}\,, \\[3mm]
\notag
	n^2  & ~~=~~ \qquad~~\, 0\,,  \\[2mm]
	     & ~~~\,\vdots          \\[2mm]
\notag
	n^k  & ~~=~~ \qquad~~\, 0\,,  \\[2mm]
\notag
 	     & ~~~\,\vdots          \\[2mm]
\notag
	n^{N-1} & ~~=~~ \qquad~~\, 0\,.                
\end{align}

	In this background, $ D^a $ taken at the edges of the worldsheet yields just the coupling constant:
\beq
	D^a \Big|^{\scriptscriptstyle +\infty}_{\scriptscriptstyle -\infty} ~~=~~    r\,.
\eeq
	We will see that the one-loop corrected contribution to the central charge of the kink is
\beq
       \mc{Z} ~~\supset~~ \frac{N}{2\pi}\, M^1\, \ln\, \frac{   |M^a|   }
                                                            {  \Lambda  }\,.
\eeq

	As for the Noether contribution, the quantization of the ``angle'' coordinate of the kink gives 
\beq
	i\, n\, M^1\,,
\eeq
	with $ q^1 ~=~ n $ an integer number.
	As for the other $ q^k $, the kink does not have fermionic zero-modes of $ \psi^k $ with $ k = 2,\, 3,\, ...\, N-1 $.
	However, we will argue that in the case of odd $ N $ there is a
	{\it non}-zero mode relevant to the problem of multiple towers that we consider (in fact, 
	the existence of this nonzero mode was noted by Dorey {\it et al.} \cite{Dorey:1999zk}).
	This mode describes a bound state of the kink and a fermion $ \psi^k $ for $ k \,=\, (N+1)/2 $.
     
%%%%%%%%%%%%%%%%%%%%%%%%%%%%%%%%%%%%%%%%%%%%%%%%%%%%%%%%%%%%%%%%%%%%%%%%%%%%%%%%%%
%              S E M I C L A S S I C A L   C A L C U L A T I O N                 %
%           O F   T H E   C E N T R A L   C H A R G E   I N   C P ^ 2            %
%%%%%%%%%%%%%%%%%%%%%%%%%%%%%%%%%%%%%%%%%%%%%%%%%%%%%%%%%%%%%%%%%%%%%%%%%%%%%%%%%%
\subsection[Semiclassical calculation of the central charge in CP$^2$]
	{Semiclassical calculation of the central charge in CP\boldmath{$^2$}}
\label{semclas}
     
	If the twisted masses $M_a$ satisfy the condition
\beq
	| M_a | ~~\gg~~ \Lambda\,,
\eeq
	then we find ourselves at weak coupling where the one-loop calculation of the
	central charges will be sufficient for our purposes. 
	This calculation can be carried out in a straightforward manner for all CP$^{N-1}$ models, 
	but for the sake of simplicity we will limit ourselves to CP$^2$. 
	Generalization to larger $N$ is quite obvious. 
	
\begin{figure}
\begin{center}
\epsfxsize=5.0cm
\epsfbox{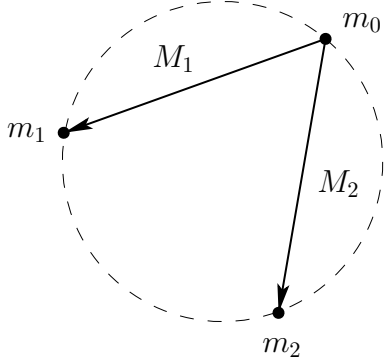}
\caption{Mass parameters $ M_1 $ and $ M_2 $ figuring in the geometric formulation of CP$^2$.}
\label{fM}
\end{center}
\end{figure}

	In CP$^2$ there are two twisted mass parameters, $M_1$ and $M_2$, as shown in Fig.~\ref{fM}.
	Accordingly, there are two U(1) charges, see Eq.~\eqref{u1cur}. 
	The Noether charges are not renormalized; 
	therefore we will focus on the topological part represented by the Killing potentials, 
	which are renormalized. 

	One-loop calculations are most easily performed using the background field method.
	For what follows it is important that $ \phi^2_\text{b} ~\equiv~ 0$ for the kink under consideration.
	If so, all off-diagonal elements of the metric $ g_{i\bj} $ vanish, while the diagonal elements take the form
\begin{align}
\nonumber
	g_{1\bar 1}^{\rm b} & ~~\equiv~~ g_{1\bar 1}\, \Big|_{\phi_{\rm b}} ~~=~~ \frac{2}{g_0^2}\,\frac {1}{\chi^2}\,,
	\\[3mm]
	g_{2\bar 2}^{\rm b} & ~~\equiv~~ 
	g_{2\bar 2}\,\Big|_{\phi_{\rm b}} ~~=~~ \frac{2}{g_0^2}\,\frac {1}{\chi}\,,
\end{align}
where
\beq
	\chi ~~=~~ 1 ~~+~~ \left|\, \phi^1_{\rm b} \,\right|^2
\eeq
	At the boundaries $\phi_{\rm b}^{1,2}$ take their (vacuum) coordinate-independent values;
	therefore, the Lagrangian for the quantum fields can be written as
\beq
	\mathcal{L}  ~~=~~  g_{1\bar 1}^{\rm b}\, \left|\, \p^\mu\, \phi^1_{\rm qu} \,\right|^2  ~~+~~
			g_{2\bar 2}^{\rm b}\, \left|\, \p^\mu\, \phi^2_{\rm qu} \,\right|^2  ~~+~~ ...
\label{324}
\eeq
	where the ellipses stand for the terms irrelevant for our calculation.

	The Killing potentials can be expanded in the same way. Under the condition $\phi^2_{\rm b} ~\equiv~ 0$ 
	we arrive at
\begin{align}
\nonumber
	D^1  & ~~=~~
 	D^1 \, \Big|_{\phi_{\rm b}}  ~~+~~  
	\frac{2}{g_0^2}\, \frac{1 \,-\, \left|\, \phi^1_{\rm b} \,\right|^2}{\chi^3}\, 
	\left|\,  \phi^1_{\rm qu}\, \right|^2 ~~-~~  
	\frac{2}{g_0^2}\, \frac{\left|\, \phi^1_{\rm b} \,\right|^2}{\chi^2}\, 
	\left|\,  \phi^2_{\rm qu} \,\right|^2  ~~+~~  ...\,,
	\\[3mm]
	D^2  & ~~=~~  0\,.
\label{325}
\end{align}
	Equation \eqref{324}
	implies that the Green's functions of the quantum fields are
\beq
	\left\langle\, \phi^1_{\rm qu}\,,~ \phi^1_{\rm qu} \,\right\rangle  ~~=~~  
	\frac{g_0^2\,\chi^2 }{2}\, \frac{i}{k^2 \,-\, \left|\, M \,\right|^2}\,,
	\qquad
	\left\langle\, \phi^2_{\rm qu}\,,~ \phi^2_{\rm qu} \,\right\rangle  ~~=~~ 
	\frac{g_0^2\,\chi }{2 }\, \frac{i}{k^2 \,-\, \left|\, M \,\right|^2}\,.
\label{326}
\eeq
	where 
\beq
	\left|\, M \,\right|  ~~\equiv~~ \left|\, M_1 \,\right|  ~~\equiv~~ \left|\, M_2 \,\right|\,.
\eeq
	Now, combining \eqref{325} and \eqref{326} to evaluate the tadpoles graphs of Fig.~\ref{ftad}
	with $\phi^1_{\rm qu}$ and $\phi^2_{\rm qu}$ running inside we arrive at
\begin{align}
\nonumber
	D^1_{\rm (one-loop)} 
	&  ~~=~~
	\frac{1}{4\pi}\, \ln \frac{\left|\, M_{\rm uv} \,\right|^2}{\left|\, M \,\right|^2}\,
	\lgr \frac{1 \,-\, \left|\, \phi^1_{\rm b} \,\right|^2}{\chi} 
		~-~ \frac{ \left|\, \phi^1_{\rm b} \,\right|^2}{\chi}
	\rgr^{\phi_{\rm b}^{1} \;=\; 0}_{\phi_{\rm b}^{1} \;=\; \infty}
	\\[3mm]
	&  ~~=~~  
	\frac{1}{4\pi}\, \ln \frac{ \left|\, M_{\rm uv} \,\right|^2}{\left|\, M \,\right|^2}
	\left(\, 2 ~+~ 1 \,\right).
\end{align}
	The first and second terms in the parentheses come from the $\phi^1_{\rm qu}$ and $\phi^2_{\rm qu}$ loops, respectively. 
	In the general case of the CP$^{N-1}$ model
	one must replace $ 2 ~+~ 1$ with $ 2 ~+~ 1 \,\times\, (N \,-\, 2) ~=~ N$.

	This information allows us to obtain the contribution of the Killing potential to the central charge at one loop, namely,
\beq
	\Delta_{\rm K}{\mathcal Z}  ~~=~~  
	-\, 2\,M_1\, \lgr 
		\frac{1}{g_0^2} ~-~
		\frac{3}{4\pi}\, \left(\, \ln\; \left| \frac{M_{\rm uv}}{M} \right| \;+\; 1
			\,\right) 
		\rgr .
\label{329}
\eeq
	Note that the renormalized coupling in the case at hand is \cite{Novikov:1984ac}
\beq
	\frac{1}{g^2}  ~~=~~  \frac{1}{g_0^2} ~~-~~
	\frac{3}{4\pi}\, \ln\; \left|\, \frac{M_{\rm uv}}{M} \,\right|  ~~\equiv~~   
	\frac{3}{4\pi}\, \ln\; \left|\, \frac{M}{\Lambda} \,\right|\,.
\label{330}
\eeq
	For the generic CP$^{N-1}$ model the coefficient 3 in front of the logarithm 
	in \eqref{330} is replaced by $ N $. 
	Equation \eqref{330} serves as a (standard) definition of the dynamical scale parameter $ \Lambda $ 
	in perturbation theory in the Pauli-Villars scheme.

	This is not the end of the story, however. We must add to $\Delta_{\rm K}{\mathcal Z}$
	a part of the central charge associated with the Noether terms in \eqref{21}, which accounts for the quantization of the
	fermion zero modes as well as effects of the $\theta$ term.
	We postpone this until Section~\ref{kinksolu}.
	At this point we are able to compare the quasi-classical central charge 
	to the weak-coupling expansion of the exact formula.

%*********************************************************************************
%                 W E A K - C O U P L I N G   E X P A N S I O N                  %
%*********************************************************************************
\subsubsection{Weak-coupling expansion}
\label{weakexp}

	The exact central charge determining the BPS spectrum is given by the difference of the values of the superpotential in  two vacua.
	As described in Section~\ref{exact}, the general formula adjusted for CP$^2$ is as follows \cite{Bolokhov:2011mp}:
\beq
       \mc{Z}\Big|^{\scriptscriptstyle +\infty}_{\scriptscriptstyle -\infty} 
	~~=~~~ 
       U_0(m_0) ~~+~~ i\, m_k\,, \qquad\qquad\qquad k ~=~ 0\,,~1\,,~2.
\eeq
	We also introduce a term $ i\, n\, M_1 $ for dyonic excitations.
	Overall, we expand
\begin{align}
\label{331}
       \mc{Z}\Big|^{\scriptscriptstyle +\infty}_{\scriptscriptstyle -\infty} 
       & 
	~~=~~~ 
       U_0(m_0)  ~~+~~ i\, n\, M_1 ~~+~~ i\, m_k
	\\[3mm]
\notag
       &
	\stackrel{|m_0|\to\infty}{\longrightarrow} 
       \frac{3}{2\pi}\, M_1\, \Big\{ \ln\, \frac {   |M_1|   }
                                                 {  \Lambda_\text{np}  } \,-\, 1 
						 ~-~ \frac{\pi}{6\sqrt{3}}
						\Big\}  ~~+~~  i\, n  \, M_1
	\\[3mm]
\notag
	&
	~~~
	~~+~~ i\,  m_k \,~~+~~ ...\,,
	\qquad\qquad\ \text{with}~ k ~=~ 0,~1,~ 2,
\end{align}
	where the ellipsis represents  suppressed terms fading out as inverse powers of the large mass parameter.
	Although for our purposes we pick $ m_0 $ to be real, expansion \eqref{331} is actually valid in 
	the whole sector of 
\beq
	 -\,\pi / N ~~<~~ \text{Arg}\;m_0 ~~<~~ +\,\pi / N \,.
\eeq
	We have explicitly re-introduced the dynamical scale $ \Lambda $, which we mark as a 
	``nonperturbative'' quantity, as being part of the exact superpotential.

	Various contributions are easy to identify in Eq.~\eqref{331}.
	The first term in the figure bracket can be identified with the running coupling constant.
	The second term comes from the anomaly.
	The third term is the real addition to the logarithm, and should be absorbed into the dynamical scale $ \Lambda $.
	This way, matching with the quasiclassical calculation of the central charge is achieved:
\beq
	\Lambda_\text{pt} ~~=~~ e^{\, \pi / 6 \sqrt{3}}\, \Lambda_\text{np}\,.
\label{lambda}
\eeq

	The second term in Eq.~\eqref{331} gives the contribution for dyonic excitations.
	Dyonic states with $ n ~ \neq~ 0 $ only exist for $ k ~=~ 0 $ or $ k ~=~ 1 $
	(the states with $ k ~=~ 1 $ differ from those with $ k ~=~ 0 $ by a shift of $ n $).

	Now taking into account relation (\ref{lambda}) we see that  the first term in the second line in  (\ref{331})
	coincides with the quasiclassical expression for the topological contribution to the central charge (\ref{329}).
	Namely,
\beq
	U_0(m_0) ~~\approx~~ \Delta_{\rm K}{\mathcal Z}\,.
\label{UapproxT}
\eeq
	Generalizing this to arbitrary $ m_0 $ we conclude that the topological contribution to the central charge 
	is exactly given by $ U_0 $, namely
\beq
	\int\, dz\, \p_z\, O \, ~~=~~ U_0(m_0),
\label{U=T}
\eeq
	see (\ref{21}).

	This means that the last two contributions in (\ref{331}) represent global U(1) charges.
	Splitting
\beq
	m_0  ~~=~~  -~ \frac13\, M_1  ~~-~~ \frac13\, M_2
\eeq
	we find that the two U(1) charges for the tower with $ k \,=\, 0 $ in (\ref{331}) are
\beq
	q^1  ~~=~~  n  ~~-~~  \frac13, \qquad\qquad q^2 ~~=~~ -\,\frac13.
\label{qcharges}
\eeq
	The appearance of fractional charges $ 1/3 $ (or $ 1/N $ for CP$^{N-1}$ theory) is in accord with 
	the results of \cite{FeIn}, 
	where $ 1/N $ fractional charges were found for a similar class of two-dimensional models.
	Also note, that a fractional half-integer U(1) charge was identified in the CP$^1$ model in \cite{SVZw}.

	Identification of $ U_0 $ with the topological part of the central charge is also natural at strong coupling.
	At small $ m_0 $ we have three states with
\begin{align}
	\mc{Z} &  ~~=~~ U_0(m_0)  ~~+~~ i\, m_k 
	\\
\notag
	& 
	~~\approx~~   -\, \frac{3}{2\pi}\, \left(\, e^{2 \pi i / 3 } \;-\; 1 \,\right) \, \Lambda
	~~+~~  i\,m_k, 
	\qquad\qquad 
	k ~=~ 0\,,~1\,,~2\,.
\end{align}
	If $ U_0 $ is identified with the topological contribution then 
	the global charge contributions come symmetrically for all three states. 
	In particular, it is known that in the limit $ m_0 ~\to~ 0 $ these states fill a fundamental representation of SU(3).

	We re-write Eq.~\eqref{331} for convenience as 
\beq
\label{weak}
	\mc{Z}  ~~=~~  \left(r -\frac{3}{2\pi} \right)\, M_1  ~~+~~  i\, n  \, M_1
	~~+~~ i\,  m_k  \,, \qquad\quad k ~=~ 0\,,~1\,,
\eeq
	where the coupling constant $ r $ is
\beq
	r  ~~=~~  \frac{3}{2\pi}\,  \ln\, \frac {   |M_1|   }
                                               {  \Lambda_\text{pt}  } \, .
\eeq

%%%%%%%%%%%%%%%%%%%%%%%%%%%%%%%%%%%%%%%%%%%%%%%%%%%%%%%%%%%%%%%%%%%%%%%%%%%%%%%%%%
%     Q U A S I C L A S S I C A L   K I N K   S O L U T I O N   I N   C P  ^ 2   %
%%%%%%%%%%%%%%%%%%%%%%%%%%%%%%%%%%%%%%%%%%%%%%%%%%%%%%%%%%%%%%%%%%%%%%%%%%%%%%%%%%
\subsection[Quasiclassical Kink Solution in CP$^2$]
	{Quasiclassical Kink Solution in CP\boldmath{$^2$}}
\label{kinksolu}

	In this section we will briefly discuss the kink solution in CP$^2$. 
	In fact, the bosonic part (and its quantization), 
	as well as the fermion zero mode in $ \psi^1 $ are the same as in CP$^1$ \cite{SYrev}. 
	The crucial difference is the occurrence of a localized (but nonzero!) mode in $ \psi^2 $.

\subsubsection{Kink solution}

	In the classical kink solution the field $ \phi^2 $ is not involved. 
	The BPS equation for $ \phi^1 $ is the same as in CP$^1$, namely,
\beq
	\p_z\, \phi ~~=~~ |\, M \,| \, \phi\,.
\label{13twentysix}
\eeq
	The solution of this equation  can be written as
\beq
	\phi\, (z)  ~~=~~  e^{ |\, M \,|\; (z \,-\, z_0) ~-~ i\,\alpha}\,.
\label{13twentyseven}
\eeq
	Here $ z_0 $ is the kink center while $ \alpha $ is an arbitrary phase related to U(1)$_1$.
	In fact, these two parameters enter only in a combination
	$ | m |\, z_0 ~+~ i\,\alpha $  ~---~  the kink center is complexified. 
	The effect of the modulus $ \alpha $ explains the occurrence of 
	$ n\, M_1 $ in the Noether part.

%*********************************************************************************
%       Q U A N T I Z A T I O N   O F   T H E   B O S O N I C   M O D U L I      %
%*********************************************************************************
\subsubsection{Quantization of the bosonic moduli}

	To carry out  conventional
	quasiclassical quantization we, as usual,
	assume the moduli\index{moduli} $ z_0 $ and $ \alpha $ in Eq. \eqref{13twentyseven}
	to be weakly time-dependent, substitute \eqref{13twentyseven}
	into the bosonic part of the Lagrangian, integrate over $ z $
	and arrive at
\beq
{\cal L}_{\rm QM}  ~~=~~  -\,M_{\rm kink}  ~~+~~  \frac{M_{\rm kink}}{2}\; \dot z_0^2 ~~+~~  
			\left\{\,
				\frac{1}{g^2\,|\, M \,|}\; \dot\alpha^2  ~-~
				\frac{\theta}{2\pi}\;\dot\alpha
			\,\right\}\,.
\label{13twentynine}
\eeq
	The first term is the classical kink mass, the second describes
	free motion of the kink along the $z$ axis.
	The term in the braces is the most interesting.
	The variable $ \alpha $ is compact.
	Its very existence is related to the exact U(1) symmetry of the model.
	The energy spectrum corresponding
	to $ \alpha $ dynamics is quantized.
	It is not difficult to see that
\beq
E_{[\alpha ]}  ~~=~~ \frac{g^2\, |\, M \,|}{4}\; q_{\rm U(1)}^2\,,
\label{13thirty}
\eeq
	where 
$ q_{\rm U(1)} $ 
	is the U(1) charge of the soliton,
\beq
	q_\text{U(1)} ~~=~~ k ~~+~~ \frac{\theta}{2\pi}\,,\qquad k ~\in~ \mc{Z}\,.
\label{13thirtyone}
\eeq
	The U(1) charge of the kink is no longer integer
	in the presence of the $\theta$ term, as it is shifted by $\theta/(2\pi )$.

%*********************************************************************************
%                            F E R M I O N S   I N                               %
%             Q U A S I C L A S S I C A L   C O N S I D E R A T I O N            %
%*********************************************************************************
\subsubsection{Fermions in quasiclassical consideration}

	First we will  focus on the 
	zero modes of $\psi^1$ in the kink background \eqref{13twentyseven}.
	The coefficients
	in front of the fermion zero modes will become (time-dependent)
	 fermion moduli, for which we are going to build
	corresponding quantum mechanics. 
	There are two such moduli, $\bar\eta$ and $\eta$.

	The equations for the fermion zero modes are
\begin{align}
	\p_z\,\psi_L^1  ~~-~~  
	\frac{2}{\chi}\, 
	\left(\, \bphi^1\, \p_z\, \phi^1 \,\right) \,\psi_L^1  ~~-~~
	i\,\frac{1 \,-\, \bphi^1\, \phi^1}{\chi}\; |\, M \,|\; e^{i\, \beta} \,\psi_R^1
	&  ~~=~~  0\,,
\nonumber
	\\[3mm]
	\p_z\,\psi_R^1  ~~-~~  
	\frac{2}{\chi}\, 
	\left(\, \bphi^1\, \p_z\, \phi^1 \,\right) \,\psi_R^1  ~~+~~  
	i\,\frac{1 \,-\, \bphi^1\, \phi^1}{\chi}\; |\, M \,|\; e^{-\, i\, \beta} \,\psi_L^1
	&  ~~=~~  0
\label{13fourtyfour}
\end{align}
	(plus similar equations for $\bpsi$; since our operator is 
	Hermitean we do not need to consider them separately).

	It is not difficult to find solution to these 
	equations, either directly, or using supersymmetry.
	Indeed, if we know the bosonic solution \eqref{13twentyseven},
	its fermionic superpartner --- and the fermion zero modes are such 
	superpartners --- is obtained from the
	bosonic one by those two supertransformations which act on
	$ \bphi $, $ \phi $ non-trivially.
	In this way we conclude that the
	functional form of the fermion zero mode 
	must coincide  with the functional form of the bosonic  
	solution \eqref{13twentyseven}. 
	Concretely,
\beq
	\lgr
		\begin{array}{c}
			\psi_R  \\  
			\psi_L
		\end{array}
	\rgr  ~~=~~ 
	\eta \,
	\left( \frac{g^2\; |\, M\,|}{2} \right)^{1/2}
	\lgr 
		\begin{array}{c}
			-\,i\,e^{-\,i\,\beta}  \\
			1
		\end{array}
	\rgr  \,e^{|\, M \,|\; (z \,-\, z_0)}
\label{13fourtyfive}
\eeq
	and
 \beq
	\lgr
		\begin{array}{c}
			\bpsi_R^1  \\
			\bpsi_L^1
		\end{array}
	\rgr  ~~=~~  
	\ov{\eta}\, \left( \frac{g^2\; |\, M \,|}{2} \right)^{1/2} 
	\lgr
		\begin{array}{c}
			i\, e^{i\, \beta}  \\
			1
		\end{array}
	\rgr  \,e^{|\, M \,|\; (z-z_0)}\,,
\label{13fourtysix}
\eeq
	where a numerical factor is introduced to ensure proper
	normalization of the quantum-mechanical Lagrangian.
	The other solution, which at large $ z $ asymptotically behaves
	as $ e^{3\, |\, M \,|\; (z \,-\, z_0)} $ must be discarded as non-normalizable.

	Now, to
	perform   quasiclassical quantization we follow the standard route:
	the moduli\index{moduli} are assumed to be   time-dependent, and we derive 
	quantum mechanics of the moduli starting from the original Lagrangian.
	Substituting the kink solution and the fermion zero modes for
	$ \psi $ one gets
\beq
	{\cal L}'_{\rm QM}  ~~=~~ i\, \ov{\eta}\, \dot\eta\,.
\label{13fourtyseven}
\eeq
	In the Hamiltonian approach the only remnants of the fermion moduli
	are the anticommutation relations
\beq
	\{\, \ov{\eta}\; \eta \,\}  ~~=~~  1\,,\qquad\quad
	\{\, \ov{\eta}\; \ov{\eta} \,\}  ~~=~~  0\,,\qquad\quad 
	\{\, \eta\; \eta \,\}  ~~=~~  0\,,
\label{13fourtyeight}
\eeq
	which tell us that the wave function is two-component
	({\it i.e.} the kink supermultiplet is two-dimensional). 
	One can implement Eq.~\eqref{13fourtyeight} by choosing {\it e.g.} 
	$ \ov{\eta} ~=~ \sigma^+ $, $ \eta=\sigma^- $.

%*********************************************************************************
%                       C O M B I N I N G   B O S O N I C                        %
%                    A N D   F E R M I O N I C   M O D U L I                     %
%*********************************************************************************
\subsubsection{Combining bosonic and fermionic moduli}
\label{cbfm}

	Quantum dynamics of the kink at hand is summarized by the
	Hamiltonian
\beq
	H_{\rm QM}  ~~=~~  \frac{M_{\rm kink}}{2}\;\dot{\ov{\zeta}}\, \dot\zeta
\label{13fourtynine}
\eeq
	acting in the space of two-component wave functions.
	The variable $ \zeta $ here is a complexified kink center,
\beq
	\zeta  ~~=~~  z_0  ~~+~~  \frac{i}{|\, M \,|}\, \alpha\,.
\label{13fifty}
\eeq
	For simplicity, we set the vacuum angle $ \theta ~=~0 $ for the time being
	(it will be reinstated later). 

	The original field theory we deal with has four conserved supercharges.
	Two of them, $ {\mathcal Q} $ and $ \ov{\mathcal Q} $, 
	act trivially in the critical kink sector. In moduli quantum
	mechanics they take the form
\beq
	{\mathcal Q}  ~~=~~  \sqrt{ M_0 }\; \dot\zeta\, \eta\,,
	\qquad\qquad
	\ov{\mathcal Q}  ~~=~~  \sqrt{ M_0 }\; \dot{\ov{\zeta}}\, \ov{\eta} \, ;
\label{13fiftyone}
\eeq
	they do indeed vanish provided that the kink is at rest.
	The superalgebra\index{superalgebra} describing kink quantum mechanics is
	$ \{\,  \ov{\mathcal Q}\, {\mathcal Q} \,\} ~=~ 2\, H_{\rm QM} $. 
	This is nothing but Witten's ${\mathcal N} = 1$
	supersymmetric quantum mechanics  (two supercharges).
	The realization we deal with is peculiar and distinct from 
	that of Witten. 
	Indeed, the standard Witten quantum mechanics
	includes one (real) bosonic degree of freedom and two fermionic, while
	we have two bosonic degrees of freedom,
	$ x_0 $ and $ \alpha $. 
	Nevertheless, the superalgebra remains the same
	due to the fact that the bosonic coordinate is complexified.
	 
	Finally, to conclude this section, let us calculate the U(1) charge
	of the kink states. Starting from the expression  for the U(1)$_1$ current we
	substitute the fermion zero modes and get\,\footnote{To set the scale properly, so that
	the U(1) charge of the vacuum state vanishes, one must
	antisymmetrize
	the fermion current, 
	$ \bpsi\, \gamma^\mu\, \psi ~\to~ (1/2) \lgr \bpsi\, \gamma^\mu\, \psi \;-\; \bpsi{}^c\, \gamma^\mu\, \psi^c \rgr $ 
	where the superscript $c$ denotes $C$ conjugation.}
\beq
	\Delta q_{\rm U(1)}  ~~=~~  \frac{1}{2}\, \left[\, \ov{\eta}\, \eta \,\right]
\label{13fiftytwo}
\eeq
	(this is to be added to the bosonic part, Eq.~\eqref{13thirtyone}).
	Given that $ \ov{\eta} ~=~ \sigma^+ $ and $ \eta ~=~ \sigma^- $ we arrive at
	$ \Delta q_{\rm U(1)} ~=~ \frac{1}{2}\, \sigma_3 $. 
	This means that the U(1)$_1$ charges of two kink states in the supermultiplet
	split from the value given in Eq.~\eqref{13thirtyone}:
	one has the U(1)$_1$ charge
$$
	k  ~~+~~  \frac{1}{2}  ~~+~~  \frac{\theta}{2\pi}\,,
$$
and another
$$
	k  ~~-~~  \frac{1}{2} ~~+~~  \frac{\theta}{2\pi}\,.
$$
	If $ \theta $ is zero, quantization of the fermionic moduli associated with the fermion zero modes
	produces a half-integer U(1) charge. 
	Previously it was argued \cite{FeIn} that in the case at hand the minimal charge
	should be $ 1/N $ ({\it e.g.} in Section~\ref{weakexp} we found 1/3 for CP$^2$ theory). 
	Apparently for $ N \;>\; 2 $ the non-zero fermion modes should also play a role 
	in the determination of the U(1) charges, but we will not dwell on this issue.

%*********************************************************************************
%                          B O U N D   S T A T E S                               %
%*********************************************************************************
\subsubsection[Bound States of $\psi^k$]
	{Bound States of \boldmath{$\psi^k$}}
\label{bound}

	Our calculations here will in fact be applicable to CP$^{N-1}$ with any $ N $ 
	and arbitrarily set-up twisted masses.
	Bound states are described by non-zero fermionic modes. 
	As we claimed, there is a localized non-zero mode in CP$^{N-1}$ theory.
	To find the non-zero mode, we write out the linearized Dirac equations in the background
	of the $ \phi^1 $ kink.
	For convenience, we rescale the variable $ z $ into a dimensionless variable $ s $:
\beq
	s ~~=~~ 2\, |M^1|\, z\,.
\eeq
	Then the kink takes the form
\beq
	\phi^1(s) ~~=~~ e^{s/2}\,,\qquad\qquad\text{and}\quad \phi^k(s) ~~=~~ 0 \qquad \text{for}~~ k ~>~ 1\,,
\eeq
	or
\begin{align}
\notag
	n^0  & ~~=~~ \frac {             1              }
                           {    \sqrt{ 1 ~+~ e^s }      }\,, \\[3mm]
\notag
	n^1  & ~~=~~ \frac {          e^{s/2}           }
                           {    \sqrt{ 1 ~+~ e^s }      }\,, \\[3mm]
\notag
	n^2  & ~~=~~ \qquad~~\, 0\,,  \\[2mm]
 	     & ~~~\,\vdots          \\[2mm]
\notag
	n^k  & ~~=~~ \qquad~~\, 0\,,  \\[2mm]
\notag
 	     & ~~~\,\vdots          \\[2mm]
\notag
	n^{N-1} & ~~=~~ \qquad~~\, 0\,.                
\end{align}

	The masses will also turn dimensionless by the same factor,
\beq
	\mu^l  ~~=~~ \frac{ m^l }
                        {2 |M^1|}\,,
	 \qquad
	 \text{and}
	 \qquad
	 \mu_G^a ~~=~~ \frac{ M^a }
                           {2 |M^1|}\,,
\eeq
	written both for gauge and geometric formulations.

	The linearized Dirac equations for the fermion $ \psi^k $ with $ k ~>~ 1 $ then look as
\begin{align}
\notag
	\Big\{ \p_s  ~-~ \frac{1}{2}\, f(s) \Big\}\,\, \psi_R^k   ~~+~~  i \lgr  \mu_G^1\, f(s)  ~-~  \mu_G^k \rgr\! \cdot \psi_L^k  
	& ~~=~~ \phantom{-} i\, \lambda\, \psi_L^k   \\[2mm]
	\Big\{ \p_s  ~-~ \frac{1}{2}\, f(s) \Big\}\,\, \psi_L^k   ~~-~~  i \lgr \ov{\mu}{}_G^1\, f(s)  ~-~ \ov{\mu}{}_G^k \rgr\! \cdot \psi_R^k 
	& ~~=~~ - i\, \ov{\lambda}\, \psi_R^k \,.
\end{align}
	Here $ f(s) $ is a real function
\beq
	f(s) ~~=~~ \frac{     e^s     }
 	                { 1  ~+~  e^s }\,.
\eeq
	Eigenvalue $ \lambda $ is zero for zero-modes, or gives the energy for non-zero modes.
	If one starts from the gauged formulation, one arrives at a simpler system, which can be obtained from the
	above one by a redefinition of the functions.
	In that system, function $ f(s) $ is absent from the first (kinetic) terms in both equations.
	That is, the conversion between the geometric and gauge formulations is precisely such as to remove the inhomogeneous term from the
	figure brackets,
\begin{align}
\notag
	\p_s\, \xi_R^k  ~~+~~  i \lgr \mu_G^1\, f(s) ~-~ \mu_G^k \rgr\! \cdot \xi_L^k  & ~~=~~ \phantom{-} i\, \lambda\, \xi_L^k \\[2mm]
	\p_s\, \xi_L^k  ~~-~~  i \lgr \ov{\mu}{}_G^1\, f(s) ~-~ \ov{\mu}{}_G^k \rgr\! \cdot \xi_R^k & ~~=~~ - i\, \ov{\lambda}\, \xi_R^k \,.
\end{align}
       
	This system does not allow normalizable zero modes.
	However, there is a normalizable non-zero mode with the energy given by the absolute value of 
\beq
       \lambda  ~~=~~  -\, \mu_G^k  ~~+~~ \cos\, \text{Arg}\,\, \frac{\mu_G^k}{\mu_G^1}  \,\cdot\, \mu_G^1\,.
\eeq
	The mode is
\begin{align}
\notag
	\xi_R^k  & ~~=~~  \lgr \frac{  e^{\alpha s}  }
	                            {   1 ~+~ e^s    }  \rgr^{ 1/2 }  \\[2mm]
	\xi_L^k  & ~~=~~  - i\, \frac{ \ov{\mu}{}_G^1 }
	                             {  | \mu_G^1 |   } \cdot \xi_R^k\,.
\end{align}
	
	It is straightforward to show that there exists a corresponding bosonic mode. 
	The relevant part of the Lagrangian is,
\beq
	\mc{L} ~~=~~ g_{i\bj}\, \p_\mu \phi^i\, \p^\mu\, \bphi{}^\bj 
	       ~~+~~ g_{i\bj}\, \text{Re}\, ( M^i\, \ov{M}{}^\bj ) \cdot \phi^i\, \bphi^\bj ~~+~~ ...
\eeq
	The corresponding equations of motion are
\begin{align}
\notag
	g^{i\bk}\, \frac{\delta \mc{L}}
                        {\delta\, \bphi{}^\bk} 
	& ~~=~~
	-~~ \p_\mu^2\, \phi^i 
	~~-~~ \Gamma^i_{pl} \cdot \p_\mu \phi^p\, \p^\mu \phi^l
	\\[1mm]
	&~~~~~~
	~~+~~
	(\, g^{i\bk}\, \p_{\bk}\, g_{l\bj} \,) \cdot \p_\mu \phi^l\, \p^\mu\, \bphi{}^\bj 
	~~-~~ (\, g^{i\bk}\, \p_\bj\, g_{l\bk} \,) \cdot \p_\mu \phi^l\, \p^\mu \bphi{}^\bj 
	\\[3mm]
\notag
	&~~~~~~
	~~+~~ (\, g^{i\bk}\, g_{l\bk} \,) \cdot \text{Re} \left(\, M_l\, \ov{M}{}_{\bk} \,\right) \cdot \phi^l
	\\[2.5mm]
\notag
	&~~~~~~
	~~+~~ (\, g^{i\bk}\, \p_\bk\, g_{l\bj} \,) \cdot 
		\text{Re} \left(\, M_l\, \ov{M}{}_{\bk} \,\right) \cdot \phi^l\, \bphi{}^\bj\,.
\end{align}
	Linearization of the equations of motion leads to the equation for the eigenmode $ \phi^k $:
\beq
	-\, \p_s^2\, \phi^k  ~~+~~ f(s) \cdot \p_s \phi^k ~~+~~ \!\! 
	\lgr \left| \mu_G^k \right| \;-\; 2\, \text{Re}\, \left( \mu_G^1\, \ov{\mu}{}_G^k \right)\, f(s) \rgr \phi^k 
	~~=~~ \lambda \cdot \phi^k\,,
\eeq
	where $ k \,>\, 1 $.
	The solution is the same as for the fermionic mode,
\beq
	\phi^k  ~~=~~  \lgr \frac{  e^{\alpha s}  }
	                         {   1 ~+~ e^s    }  \rgr^{ 1/2 }\,,
	\qquad\qquad
	\alpha ~=~ \text{Re}\; \frac{M_k}{M_1}\,,
\eeq
	with the corresponding eigenvalue
\beq
\label{E}
	E^2 ~~=~~ \left( 2\, | M_1 | \right)^2\, \lambda ~~=~~ \Big|\, M_k\, \sin \text{Arg}\; \frac{ M_k }{ M_1 } \,\Big|^2\,.
\eeq
	As we said, in this form the eigenmodes are actually correct for CP$^{N-1}$ with arbitrary twisted masses,
	not necessarily with ones sitting on the circle.

	That these modes are BPS can be seen from the expansion of the central charge
\beq
\label{asympt}
       |\, r \cdot M_1  ~+~ i\, M_k \,|  ~~=~~ r \cdot | M_1 |  ~~-~~ | M_k | \cdot \sin \text{Arg}\; \frac { M_k } 
                                                                                                            { M_1 }  
                                                                ~~+~~ ... \,,
\eeq
	in the large coupling constant $ r $.
	This is the central charge of the bound state of a fermion and the kink 
	as discovered by Dorey {\it et al.} \cite{Dorey:1999zk}, written semi-classically.

	The normalizability of the eigenmodes translates into
\beq
\label{norm}
	0 ~~<~~ \text{Re}\; \frac{ M_k }
                                 { M_1 } ~~<~~ 1\,.
\eeq
	In the case of $ \mc{Z}_N $ symmetric masses $ m_k $, this condition is only met for $ k \,=\, (N+1)/2 $
	when $ N $ is odd, and never met when $ N $ is even.

\newpage
%%%%%%%%%%%%%%%%%%%%%%%%%%%%%%%%%%%%%%%%%%%%%%%%%%%%%%%%%%%%%%%%%%%%%%%%%%%%%%%%%%
%                                                                                %
%                               S P E C T R U M                                  %
%                                                                                %
%%%%%%%%%%%%%%%%%%%%%%%%%%%%%%%%%%%%%%%%%%%%%%%%%%%%%%%%%%%%%%%%%%%%%%%%%%%%%%%%%%
\section{The Spectrum and CMS}
\label{secspectrum}
\setcounter{equation}{0}

	In this section we will focus on the
	evolution of the BPS spectrum from weak to strong coupling (i.e. $|m_0|$ changing from infinity to zero),
	and in the opposite direction.
	We will discuss in detail how in CP$^2$ theory the weak coupling  
	spectrum turns into the strong coupling	one, as $ m_0 $ is taken 
	from the periphery towards the center of the $ m_0 $ complex plane.
	The inverse way will of course involve the corresponding inverse processes, as well as a decay of a 
	strong coupling state.

	As was indicated earlier, in the vicinity of the center of the complex plane there are three states
	whose masses correspond to the absolute values of\,\footnote{We define our nomenclature by tagging every BPS state with its central charge}
\beq
\label{sstates}
	\W_1 ~~-~~ \W_0   ~~+~~ i\, m_k\,,   \qquad\qquad k ~=~ 0\,,~1\,,~2\,.
\eeq
	If one uses Eq. (\ref{sstates}) and  naively extrapolates this formula to calculate  the  masses at large $ |m_0| $, e.g. through Eq.~\eqref{weak},
	one  finds that the lightest state is $ \W_1 ~-~ \W_0 ~+~ i\, m_2 $.
	However, as we will show shortly, this state $ \W_1 ~-~ \W_0 ~+~ i\, m_2 $ hits a CMS at a certain domain at strong coupling
	and decays. 
	Therefore, it does not exist at large  $ |m_0| $, which is a blessing, because otherwise
	its mass would vanish somewhere between the strong and week coupling domains, an impossible event.
	
	Among the two other  states, the lightest is
\beq
\label{monopole}
	\M ~~=~~ \W_1 ~~-~~ \W_0   ~~+~~ i\, m_0\,,
\eeq
	which is natural to identify with the monopole $ ({\mathcal M}) $ of the four-dimensional theory.
	If we use Eq.~\eqref{weak} (a valid procedure in this case), to extrapolate to
	large $ |m_0| $, we will see that this state lies at the bottom of one of the two towers existing 
	in the quasiclassical domain, namely of the tower that had been originally identified in \cite{Dorey:1998yh}.
	Let us call it the $A$ tower.  
	It has a unit topological charge, while the charges with respect to the two U(1)'s  are fractional, see (\ref{qcharges}).

	Note that Eq.~\eqref{monopole} precisely agrees with Eq.~\eqref{ZM} if the branch cuts are taken into account.
	This justifies our identification of the monopole in Eq.~\eqref{ZM} and choosing AD$^{(0)}$ as the reference point.
	We will return to this discussion below.
	
	The remaining third state in (\ref{sstates}), with $k=1$, will be a dyonic state belonging to the same
	$A$ tower, with a larger U(1) charge and thus, heavier at weak coupling.

	If we introduce the quantities
\beq
\label{quark}
	\Q_{ik} ~~=~~ i\, \left( m_i  ~-~ m_k \right)\,,
\eeq
	which will later denote the central charge of elementary quanta, 
	the  two states 
	$ \W_1 ~-~ \W_0 ~+~ i\, m_{1,2} $
	of the strong coupling domain will have the quantum numbers 
\beq
\label{twostates}
	\D  ~~=~~  \M ~~+~~ \Q_{10}  \qquad\qquad  \text{and}  \qquad\qquad   
	\V^2  ~~=~~  \M ~~+~~ \Q_{20}
\eeq
(upon extrapolation to weak coupling).
	The first state will be identified as the dyon in four-dimensions. It is tempting to say that
	the second state in (\ref{twostates}) lies at the bottom of the second tower which will be referred to as the
	$ B $ tower. 
	However, this is not the case. 
	In fact, we were preoccupied with this scenario for some time.
	The continuation of $ \V^2 $ to weak coupling 
	does not exist. 
	The second tower $ B $ detected at weak coupling \cite{Bolokhov:2011mp},
	is not analytically connected with $ \V^2 $.
	When $ |m_0| $ decreases, the $ B $ tower encounters multiple CMS and disappears
	before reaching the strong coupling domain.
	Similarly, the state $ \V^2 $ decays on the way from strong to weak coupling.
	
	For convenience, we will use the same four-dimensional monopole-dyon nomenclature in two dimensions as well.
	Thus, we call the kink with the minimal U(1) charge a monopole $ \M $, 
	the state $ \Q_{ik} $ a quark and the state $ \D $ a dyon.

	If one takes the states \eqref{sstates} from the middle of the complex plane to the different AD points 
	(see Fig.~\ref{fcp2}), each of them will become massless at precisely one of these points.
	Namely, for each state in Eq.~\eqref{sstates}, 
	the branch cuts turn the right hand side of Eq.~\eqref{sstates} into 
	the difference $ \W_1 ~-~ \W_0 $ as one passes from the center of the plane to the corresponding point AD$^{(k)}$.
	The ``monopole'' state $ \M $ becomes massless at the point AD$^{(0)}$
\beq
	m_\text{AD}^{(0)} ~~=~~ e^{i \pi / 3}\,.
\eeq
	We conclude that we should choose precisely this AD point as a reference point, 
	in the sense described in Section~\ref{exact}.
	We make this choice because we want the mass of the lightest state to vanish at the reference point. 
	Its central charge in the area near this AD point is given by the difference of the superpotentials.
	%and thus this state has the quantum numbers
%\beq
%\label{Mtop}
%	\vec{T} ~~=~~  (\; -1\,, ~~~ 1\,, ~~~ 0 \;)\,,	\qquad\qquad
%	\vec{S} ~~=~~  (\;  0\,, ~~~ 0\,, ~~~ 0 \;)\,.
%\eeq
%	Here $ \vec{T} $ and $ \vec{S} $ stand for the topological and U(1) quantum numbers, respectively.
%	We now can see that the monopole indeed has only a topological charge. 

	As we have noticed, the logarithm branch cuts give the monopole $ \M $ different expressions in terms
	of $ \W_k $ in different parts of the physical area, see {\it e.g.} Eqs.~\eqref{ZM} and \eqref{monopole}.
	Following \cite{Bolokhov:2011mp} we introduce an analytical continuation of $ \M $ from AD$^{(0)}$ into
	the whole physical area,
\beq
\label{Mcont}
	\M  ~~=~~  \left(\, e^{2 \pi i / 3} ~-~ 1 \,\right) \cdot \W_0  ~~+~~  i\, m_0\,.
\eeq
	The function on the right hand side is continuous as $ \W_0 $ does not have branch cuts in the physical 
	region, see Fig.~\ref{fcp2}.

	We can now proceed to the classification of the states.
	So far we considered the elementary kinks interpolating between
	the vacua ({\sc \small 0}) and ({\sc \small 1}).
	The spectrum of the other kinks which interpolate between the vacua ({\sc \small 0}), ({\sc \small 1}) and ({\sc \small 2})
	is identical due to $ \mc{Z}_3 $ symmetry.
	Although the masses of these other kinks will match our ({\sc \small 0}) $ \rightarrow $ ({\sc \small 1}) spectrum, 
	their central charges will be different as complex numbers.
	This circumstance will appear important below.
	For this reason, we now need to expand our analysis and include all possible elementary kinks
\beq
	\M_{10}\,,   \qquad\quad     \M_{21}\,,     \qquad\quad\text{and}\qquad\quad   \M_{02}\,.
\eeq
	Here
\beq
	\M_{ik} \,:  ~~  ({ \scriptstyle k }) ~~\longrightarrow~~ ({ \scriptstyle i })
\eeq
	interpolates between the neighbouring vacua ($ \scriptstyle i $) and ($ \scriptstyle k $).
	The central charge of the state $ \M_{10} $ is given by Eq. \eqref{Mcont}, while those
	of the other two kinks just differ by a phase, see Eq.~\eqref{Wdiff},
\beq
	\M_{21} ~~=~~ e^{2 \pi i / 3}\, \M_{10}  
	\qquad\quad \text{and} \qquad\quad 
	\M_{02} ~~=~~ e^{- 2 \pi i / 3}\, \M_{10}\,.
\eeq
	The same relations are correct for the two states of Eq.~\eqref{twostates} as well.

	The strong coupling spectrum can then be summarized as
\beq
\label{sspectrum}
	\begin{array}{ccc}
		\M_{10} \qquad\quad    &    \D_{10} \qquad\quad    &    \V_{10}^2    \phantom{~~~.}    \\[2mm]
		\M_{21} \qquad\quad    &    \D_{21} \qquad\quad    &    \V_{21}^0    \phantom{~~~.}    \\[2mm]
		\M_{02} \qquad\quad    &    \D_{02} \qquad\quad    &    \V_{02}^1    ~~~. 
	\end{array}
\eeq
	Here, 
\beq
	\V_{21}^0  ~~=~~  e^{2 \pi i / 3} \cdot \V_{10}^2 
	\qquad\qquad  \text{and}  \qquad\qquad
	\V_{02}^1  ~~=~~  e^{- 2 \pi i / 3} \cdot \V_{10}^2 \,,
\eeq
	where $ \V_{10}^2 $ is given in Eq.~\eqref{twostates}.

	The weak coupling spectrum includes monopoles $ \M_{ik} $, elementary quanta $ \Q_{ik} $ 
	as well as dyons $ \D^{(n)}_{ik} $
\beq
	\D^{(n)}_{ik} ~~=~~ \M_{ik} ~~+~~ i\, n\, ( m_i ~-~ m_k ) \,,
\eeq
	which is just a generalization of Eq.~\eqref{twostates}.
	Their masses at weak coupling are given by Eq.~\eqref{weak}.
	Let us check that  this is the case.
	
	For $ k \,=\, 0 $ and $ n \,=\, 0 $, Eq.~\eqref{weak} provides the quasiclassical approximation for the monopole mass,
%\marginpar{\tiny where is the term $-iM_1/2$?}
\beq
	| \M_{10} |  ~~=~~  \left|\left(r -\frac{3}{2\pi} \right)\, M_1 
	~~-~~ \frac{i}{3}\,  M_1 ~~-~~ \frac{i}{3}\,  M_2 \right|\,.
\eeq
	For $ k \,=\, 0 $ and $ n \,\neq\, 0 $ one obtains the mass for the dyons,
\beq
	\left|\, \D_{10}^{(n)} \,\right|  ~~=~~  \left|\left(r -\frac{3}{2\pi} \right)\, M_1  
		~~+~~  i\, \Big( n \,-\, \frac13 \Big)  \, M_1
		~~-~~ \frac{i}{3}\,  M_2  \right| \,.
\eeq
	In the mass, the extra 1/3 are not captured at the quasiclassical level of precision, 
	as the corresponding contributions are suppressed by two powers of $ r $ relative to the main term. 
	However, they are within one-loop accuracy of the central charge calculation.
	Next, putting $ k \,=\, 1 $ is equivalent to setting $ n \,\to\, n \,+\, 1 $.
	Finally, equation \eqref{weak} is not applicable for $ k \,=\, 2 $.

	We found in Section~\ref{qclassics} that at weak coupling there are
	bound states of the ``monopole'' and the elementary quanta $ \psi^k $, with $ k \,>\, 1 $,
	as well as the dyonic excitations of such bound states.
	In the present case of CP$^2$, there is only one appropriate fermion, namely, $ \psi^2 $. 
	Although the state $ \V_{10}^2 ~=~ \M_{10} ~+~ \Q_{20} $ has precisely the quantum numbers
	of a bound state of $ \M_{10} $ and $ \psi^2 $, it is actually not.
	The primary reason is that $ \V_{10}^2 $ decays and does not exist at weak coupling as we will see below.
	If it did, its mass would be \emph{lighter} than that of the monopole, as it is easy to see.
	Indeed, the mass of such a state in the quasiclassical limit would be described by a limit of Eq.~\eqref{weak}
	given in Eq.~\eqref{asympt}
\beq
       |\, r \cdot M_1  ~+~ \Q_{20} \,|  ~~=~~ r \cdot | M_1 |  ~~-~~ | m_2 \,-\, m_0 | \cdot \sin \text{Arg}\; \frac { m_2 \,-\, m_0 } 
                                                                                                                      { m_1 \,-\, m_0 }  
                                                                ~~+~~ ... \,,
\eeq
	where the second term on the right hand side is positive. 
	Clearly, this situation is impossible, and such a bound state could not exist. 
	At the same time, the bound state of the
	monopole and the elementary quantum $ \psi^2$ certainly exists.

	The quasiclassical expression for the mass of the bound state is found using the eigenvalue from Eq.~\eqref{E},
\beq
\label{kinkpsi}
	M_\text{kink}^{\psi^2}  ~~=~~ \left|\, \M_{10} \,\right|  ~~+~~  E\,,  
	\qquad\qquad
	E  ~~=~~  \frac{\sqrt{3}}{2} \cdot | m_2 \,-\, m_0 |\,.
\eeq
	In fact, this is the mass of a bound state of the monopole and an anti-quantum of the 
	elementary excitation
$ \Q_{02} =  -\, \Q_{20} $,
	as can be checked from the expansion
\beq
	\left|\, \M_{10} ~+~ \Q_{02} \,\right|  ~~\approx~~  \left|\, r \cdot M_1 ~+~ \Q_{02} \,\right|
					        ~~\approx~~  \left|\, \M_{10} \,\right|  ~~+~~  \frac{\sqrt{3}}{2} \cdot | m_2 \,-\, m_0 |\,.
\eeq
	Such a state will not be lighter than the monopole.
	Therefore, at weak coupling the spectrum includes the bound states
\beq
	\M_{10}\,\Q_{02}  ~~=~~  \M_{10}  ~~+~~  \Q_{02}\,,
\eeq
	as well as their dyonic excitations $ \D_{10}^{(n)}\,\Q_{02} $.\footnote{The first and the second U(1) charges 
	for this state are -\,1/3 and -\,4/3.}

	Then at weak coupling the spectrum can be summarized as,
\beq
\label{wspectrum}
	\begin{array}{ccc}
		\Q_{10} \qquad\qquad    &    \D_{10}^{(n)} \qquad\qquad    &    \D_{10}^{(n)}\, \Q_{02}   \phantom{~~~.}    \\[2mm]
		\Q_{21} \qquad\qquad    &    \D_{21}^{(n)} \qquad\qquad    &    \D_{21}^{(n)}\, \Q_{10}   \phantom{~~~.}    \\[2mm]
		\Q_{02} \qquad\qquad    &    \D_{02}^{(n)} \qquad\qquad    &    \D_{02}^{(n)}\, \Q_{21}   ~~~.
	\end{array}
\eeq

	The spectrum \eqref{wspectrum} must change to \eqref{sspectrum} on the way from weak coupling to strong coupling.
	In particular, the elementary quanta, dyons with $ n ~\neq~ 0 $ and $ 1 $, and all dyonic bound states  
	must decay on their decay curves.

\begin{figure}
\begin{center}
\epsfxsize=7.5cm
\epsfbox{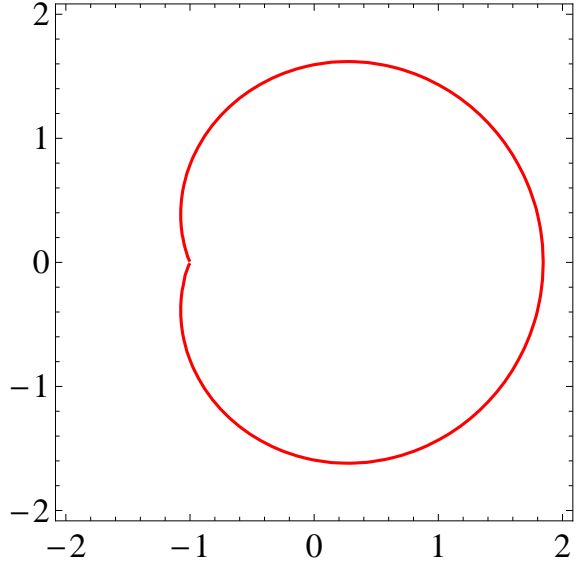}
\caption{\small The primary decay curve, in the plane of $ m^3 $}
\label{fprim}
\end{center}
\end{figure}

	Let us start from the elementary quanta.
	They decay via the reaction
\beq
\label{Qdecay}
	\Q_{ik}  ~~\longrightarrow~~  \M_{ik}\, \Q_{ik} ~~+~~ \ov{\M}{}_{ik}\,.
\eeq
	This condition, stated algebraically, means that the central charge of the ``monopole'' 
	is parallel to that of the elementary quantum,
\beq
\label{parallel}
	\M_{ik}  ~~\parallel~~  \Q_{ik}\,,
\eeq
	and this way gives us the usual ``primary'' CMS curve which passes through the AD point, see Fig.~\ref{fprim}.
	The dyonic state on the right hand side of Eq.~\eqref{Qdecay} is the dyon $ \D_{ik}^{(1)} $.
	In the region of the curve in Fig.~\ref{fprim}, the elementary quantum is heavier than both the monopole and the 
	``first'' dyon.
	These latter states are the lightest, they do not decay and become the strong coupling
	states $ k ~=~ 0 $ and $ k ~=~ 1 $ of Eq.~\eqref{sstates}.
	
	All other dyons $ D_{ik}^{(n)} $ with $ n ~\neq~ 0 $ or $ 1 $ are heavier than the monopole and quarks and can decay in the mode
\beq
\label{Ddecay}
	\D_{ik}^{(n)}  ~~\longrightarrow~~ \M_{ik} ~~+~~ n\, \Q_{ik} \,, \qquad\qquad n ~\neq~ 0\, ~~\text{or}~~ 1\,,
\eeq
	with a subsequent decay of the elementary quanta via \eqref{Qdecay}.
	Algebraically, the mode \eqref{Ddecay} gives us the same condition as \eqref{parallel}, and 
	we conclude that the elementary quanta and the dyons with $ n ~\neq~ 0 $ or $ 1 $ all decay 
	on the usual ``primary'' curve.

\begin{figure}
\begin{center}
\epsfxsize=7.5cm
\epsfbox{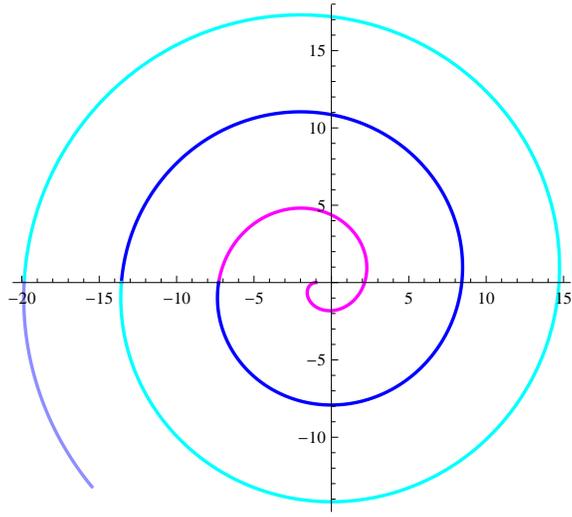}
\caption{\small The decay curve for the states $ \D^{(n)}\,\Q $ with $ n ~>~ 1 $, logarithmic plot, artificial units. 
		The spiral starts at the AD point,
		each successive curl corresponds to decay of a higher dyon}
\label{flinspiral}
\end{center}
\end{figure}

	As for the bound states, the decay of the states $ \D^{(n)}\, Q $ with positive $ n ~>~ 1 $ occurs via
\beq
\label{spirpos}
	\D^{(n)}_{10}\, \Q_{02}  ~~\longrightarrow~~  \D^{(n-1)}_{10}  ~~+~~  Q_{12}\,,
	\qquad\qquad
	n ~>~ 1\,.
\eeq
	and the cyclic permutations for the other two $ \mc{Z}_3 $ sectors.
	The corresponding curve forms a spiral that starts out from the AD point and winds outwards counter-clockwise direction.
	This occurs not to be a closed curve!
	As one makes one coil following the spiral around the complex plane, a monodromy equal to one ``unit'' of U(1)
	charge is gained,
\beq
	\W_i  ~~-~~ \W_k  ~~\longrightarrow~~  \W_i  ~~-~~ \W_k  ~~+~~  i\, ( m_i ~-~ m_k )\,.
\eeq
	This shifts all dyons up by one unit 
\beq
	\D^{(n)}_{ik}  ~~\longrightarrow~~  \D^{(n+1)}_{ik}\,.
\eeq
	Therefore, the next coil of the spiral is the decay line of the higher dyon. 
	Going further on, the successive coils of the spiral correspond to the dyons with higher and higher U(1) numbers.
	Ultimately, when one steps out from the center of the plane far enough, one comes to weak coupling,
	where ideally the bound states, as seen in quasiclassics, are stable and could not decay.
	The quasiclassical treatment assumes that the kink mass is dominated by a large logarithm,
	with everything else suppressed by powers of the coupling constant
\beq
	|\, r\, ( m_i ~-~ m_k ) ~~+~~ Q \, |  ~~=~~  r \, \lgr\, ( m_i ~-~ m_k ) ~~+~~ \frac{ E }{r} ~~+~~ ... \,\rgr .
\eeq
	Here $ E $ is given in Eqs.~\eqref{E} and \eqref{kinkpsi} and includes the ``binding energy''.
	For arbitrary large $ m_0 $, there however always exists such $ n $, that the dyonic contribution to the central charge
	is comparable to the large logarithm $ n ~\sim~ r $, 
\beq
	\mc{Z} ~~=~~ r \, ( m_i ~-~ m_k ) ~~+~~ Q ~~+~~ i\, n\, ( m_i ~-~ m_k )\,,
\eeq
	and such an expansion cannot be performed.
	Quasi-classical approximation is only applicable when
\beq
	n  ~~\ll~~  \log\; |\, m_0 \,| \,.
\eeq
	We can see that for the corresponding dyonic states the quasiclassical treatment does not apply on the spiral curve, 
	and they can decay.
	From the condition $ n ~\sim~ \log\, | m_0 | $ one immediately finds that the spiral expands exponentially 
	with the dyon number $ n $.
	In a logarithmic picture, the spiral has a regular appearance winding out linearly with $ n $, see Fig.~\ref{flinspiral}.

	The decay of bound states $ \D^{(n)}\,\Q $ with $ n ~\leq~ 0 $ occurs via the reaction
\beq
	\D_{10}^{(n)}\,\Q_{02}  ~~\longrightarrow~~  \D_{10}^{(n)}  ~~+~~  \Q_{02}\,.
\eeq
	The decay curves form the identical spiral, reflected around the real axis.
	That is, this spiral winds clockwise.

\begin{figure}
\begin{center}
\epsfxsize=7.5cm
\epsfbox{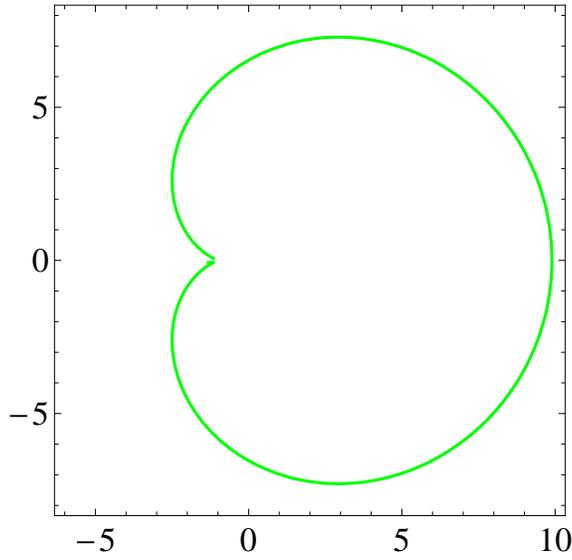}
\caption{\small The decay curve for the bound state $ \D_{10}^{(1)}\,\Q_{02} $}
\label{fddecay}
\end{center}
\end{figure}

	The two spirals account for decays of all bound states but $ \D^{(1)}\,\Q $.
	This state does not decay on the spiral.
	Instead, it has a ``topological'' decay mode
\beq
\label{DQdecay}
	\D_{10}^{(1)}\,\Q_{02}  ~~\longrightarrow~~  \bM_{02}  ~~+~~  \ov{\D}{}_{21}^{(1)}\,.
\eeq
	Here we use bars to denote the anti-states.
	The two states on the right hand side are two light states at strong coupling
	--- a monopole and a dyon --- taken from different topological sectors.
	The curve corresponding to \eqref{DQdecay} is shown in Fig.~\ref{fddecay}.

\begin{figure}
\begin{center}
\epsfxsize=7.5cm
\epsfbox{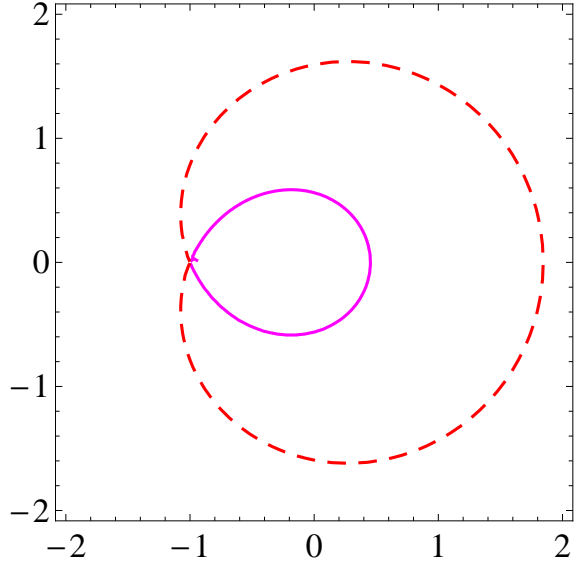}
\caption{\small The decay curve for the strong coupling state $ \V^2 $ complements the primary curve
		which is shown with a dashed line (picture in $ m^3 $ plane)}
\label{fvdecay}
\end{center}
\end{figure}

	We have found all decay curves for the weak-coupling states which are not part of the strong coupling spectrum.
	Now, going the other way, at strong coupling there is an extra state $ \V_{10}^2 $ which, 
	as we argued, cannot be identified with any bound state at the weak coupling.
	Indeed, this state does not make it to weak coupling and decays via
\beq
\label{Vdecay}
	\V_{10}^2  ~~\longrightarrow~~  \ov{\D}{}_{02}^{(1)}  ~~+~~  \bM_{21}\,.
\eeq
	Its decay curve is shown in Fig.~\ref{fvdecay} and complements the primary curve by forming an inner loop.
	Note that the decay modes \eqref{DQdecay} and \eqref{Vdecay} are precisely the reason why we had to include all
	three $ \mc{Z}_3 $ topological sectors and introduce the lower indices for the kinks $ \M_{ik} $.

\begin{figure}
\begin{center}
\epsfxsize=8.0cm
\epsfbox{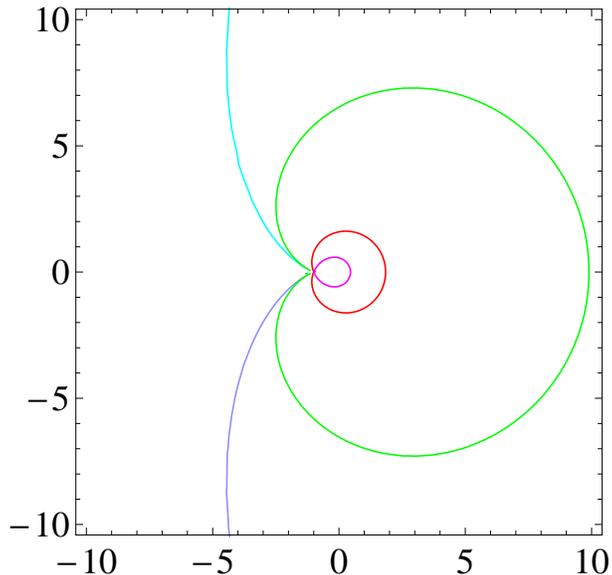}
\caption{\small The decay curves of CP$^2$ in $ m_0^3 $ plane. 
		The primary curve is shown in red.
		The two vertical whiskers are the initial coils of the two spirals}
\label{fdecays}
\end{center}
\end{figure}

	We present a collective picture of the decay curves in Fig.~\ref{fdecays}, which includes the spirals, the 
	primary curve with an inner loop, and a curve for the decay of $ \D\,\Q^{(1)} $.
	The strong coupling spectrum \eqref{sstates} is recovered in the middle of the inner-most curve.	

\newpage

%%%%%%%%%%%%%%%%%%%%%%%%%%%%%%%%%%%%%%%%%%%%%%%%%%%%%%%%%%%%%%%%%%%%%%%%%%%%%%%%%%
%                                                                                %
%                            C O N C L U S I O N                                 %
%                                                                                %
%%%%%%%%%%%%%%%%%%%%%%%%%%%%%%%%%%%%%%%%%%%%%%%%%%%%%%%%%%%%%%%%%%%%%%%%%%%%%%%%%%
\section{Conclusion}
\label{conclu}
\setcounter{equation}{0}

	We have re-examined the problem of the BPS spectrum in CP$^{N-1}$ theory with twisted masses.
	Our previous analysis \cite{Bolokhov:2011mp} revealed the existence of extra states which were not discussed 
	before.
	It was not fully clear, however, how the states seen at strong coupling were related to those detected 
	in the weak coupling domain. Now we trace their relationship in a detailed way. 
	
	First we confirmed that at weak coupling, out of $N$ possible towers, only two
	(odd $N$) or one (even $N$) are realized dynamically. 
	More exactly this statement can be formulated as follows. A quasiclassical analysis of 
	the kink bound states  shows that in our set-up
	with the $ \mc{Z}_N $ symmetric masses there is only one extra tower of states which was not known to exist
	in the case of odd $ N $, and no towers for even $ N $.
	
	For $N=3$ we identified two towers at weak coupling and clarified their dynamical origin. 
	Three states existing at strong coupling evolve (as we move towards the weak coupling) as follows. 
	Two states form the foundation of the tower $ A $. 
	The third, ``extra" state encounters a CMS and decays before reaching the weak coupling region.
	Although we did not perform a thorough analysis for arbitrary $ N $, we expect 
	that in the general case the $ N \,-\, 2 $ extra states do not reach weak coupling as well.
	Indeed, for all even $ N $ it is argued that there are no bound states they could be identified with.
	It is reasonable to assume that the extra states decay for CP$^{N-1}$ with arbitrary $ N $.

	In CP$^2$ theory, the tower $ B $ does not decay on a single curve, as was the case in CP$^1$, but rather, 
	each excited state of the tower decays on its designated branch of a spiral curve. 
	The elementary states, which correspond to quarks/gauge bosons in four dimensions, decay on the primary CMS.
	We expect a similar picture to take place for CP$^{N-1}$ with a general odd $ N $.
	Even more generally, the condition \eqref{norm} is not as restricting in a theory with arbitrary distributed 
	twisted masses.
	Multiple towers of bound states may exist at weak coupling, both for even and odd $ N $.
	Speaking of the wall crossings in terms of the decay curves becomes impractical, however.

	The correspondence between the spectra of the two-dimensional sigma model and four-dimensional U$(N)$ QCD
	in the $r=N$  Higgs vacuum elevates this picture to four dimensions.
	The bound states of kinks and elementary quanta which exist for odd $ N $ correspond to 
	bound states of dyons and quarks.
	The extra states at strong coupling decay before reaching the weak coupling domain and 
	cannot be described quasiclassically.

%%%%%%%%%%%%%%%%%%%%%%%%%%%%%%%%%%%%%%%%%%%%%%%%%%%%%%%%%%%%%%%%%%%%%%%%%%%%%%%%%%
%                                                                                %
%                        A C K N O W L E D G M E N T S                           %
%                                                                                %
%%%%%%%%%%%%%%%%%%%%%%%%%%%%%%%%%%%%%%%%%%%%%%%%%%%%%%%%%%%%%%%%%%%%%%%%%%%%%%%%%%
\section*{Acknowledgments}
\addcontentsline{toc}{section}{Acknowledgments}
We are grateful to N. Dorey and K. Petunin for making the preliminary version of their paper \cite{ndkp}
available to us.
The work of PAB and MS was supported by the DOE grant DE-FG02-94ER40823.
The work of AY was  supported
by  FTPI, University of Minnesota
and by the Russian State Grant for
Scientific Schools RSGSS-65751.2010.2.

\end{document}